\documentclass[twocolumn]{aastex631}
\usepackage{graphicx}	
\usepackage{amsmath}
\usepackage{amssymb}	

\usepackage{textgreek}

\usepackage[T1]{fontenc}

\newcommand{\p}{\partial}

\newcommand{\OmK}{\Omega_\text{K}}
\newcommand{\ts}{t_\mathrm{s}}
\newcommand{\cs}{c_\mathrm{s}}
\newcommand{\ky}{k_\mathrm{y}}

\newcommand{\rg}{\rho_\mathrm{g}}
\newcommand{\rd}{\rho_\mathrm{d}}

\newcommand{\vgx}{v_\mathrm{gx}}
\newcommand{\vgy}{v_\mathrm{gy}}
\newcommand{\vdx}{v_\mathrm{dx}}
\newcommand{\vdy}{v_\mathrm{dy}}

\newcommand{\vg}{\mathbf{v}_\mathrm{g}}
\newcommand{\vd}{\mathbf{v}_\mathrm{d}}

\newcommand{\nn}{\nonumber}
\newcommand\bb[1]{\boldsymbol{\mathbf{#1}}}

\defcitealias{dullemond+18}{D+18}
\defcitealias{lb23}{LB23}

\shorttitle{}
\shortauthors{Cui et al.}
\begin{document}

\title{Dust Growth in ALMA Rings: II. Dusty Rossby Wave Instability}

\author{Can Cui}
\altaffiliation{\href{mailto:ccui@nju.edu.cn
}{ccui@nju.edu.cn}}
\affiliation{Institute for Deep Space Exploration, Nanjing University, Suzhou 215163, China}
\affiliation{Department of Astronomy and Astrophysics, University of Toronto, Toronto, ON M5S 3H4, Canada}

\author{Konstantin Gerbig}
\affiliation{Department of Astronomy, Yale University, New Haven, CT 06511, USA}

\author{Ya-Ping Li}
\affiliation{Shanghai Astronomical Observatory, Chinese Academy of Sciences, Shanghai 200030, China}

\author{Ziyan Xu}
\affiliation{Center for Star and Planet Formation, GLOBE Institute, University of Copenhagen, Øster Voldgade 5–7, 1350 Copenhagen, Denmark}

\author{Rixin Li}
\affiliation{Department of Astronomy, Theoretical Astrophysics Center, and Center for Integrative Planetary Science, \\ University of California Berkeley, Berkeley, CA 94720-3411, USA}

\author{Cong Yu}
\affiliation{School of Physics and Astronomy, Sun Yat-Sen University, Zhuhai 519082, China}

\author{Min-Kai Lin}
\affiliation{Institute of Astronomy and Astrophysics, Academia Sinica,Taipei 10617, Taiwan, R.O.C.}
\affiliation{Physics Division, National Center for Theoretical Sciences, Taipei 10617, Taiwan, R.O.C.}

\author{Feng Yuan}
\affiliation{Center for Astronomy and Astrophysics and Department of Physics, Fudan University, 200438, Shanghai, China}

\begin{abstract}

Annular substructures serve as ideal venues for planetesimal formation. In this series, we investigate the linear stage of dust growth within rings.  The first paper examines the global streaming instability, while this study focuses on the dusty Rossby wave instability (DRWI). We perform a linear analysis of the two-fluid equations on a background pressure bump, representing annular substructures. The spectral code \textsc{Dedalus} is used to solve the linear eigenvalue problem. We identify two distinct DRWI modes: Type I, which originates from dust-modified gas RWI, and Type II, which results from dust-gas coupling. Type I and Type II modes never coexist for a given azimuthal wavenumber $\ky$, but transition between each other as $\ky$ varies. Type I modes are driven by the advection of background vorticity, whereas Type II modes involve two primary waves: Rossby waves, driven by advection, and thin waves, driven by dust-gas drag. Finally, we assess the relevance of DRWI in ALMA rings using DSHARP sources. Our findings suggest that Type I modes could explain the absence of azimuthal asymmetries in many ALMA disks, whereas Type II modes are entirely absent in all eight observed rings, implying that unresolved narrow rings or alternative mechanisms may play a role in dust growth within annular substructures.

\end{abstract}


%%%%%%%%%%%%%%%%%%%%%%%%%%%%%%%%%%%%%%%%%%%%%%%%%%
\section{Introduction}\label{in}

Protoplanetary disks are the birth places of planets, where dust particles coagulate and collapse into planetesimals, seeding further growth into terrestrial planets and cores of gas giants. Planet formation requires the growth from micron-sized dust into $\sim 10^3-10^5$ km-sized planets within several million years \citep{simon+22,Joanna+23}. In the early stage of planet formation, laboratory experiments have showed that grains of $0.1-1$ $\mu m$ in size generally stick and grow upon mutual collisions \citep{bw08,gb15,blum18}. The outcome of sticking can yield cm-sized grains at 1 AU \citep{zsom+10}, and mm-sized grains at $\sim15-50$ AU \citep{lorek+18}. 
The intermediate stage of planet formation is the conversion of mm- to cm-sized grains, or pebbles, to km-sized planetesimals. However, the growth by sticking is halted by the bouncing or fragmentation barrier in the inner part of the disk \citep{guttler+10,zsom+10}, and by the fast radial drift barrier in the outer part of the disk \citep{weidenschilling77,birnstiel_etal10,birnstiel24}. Observations of dust continuum emission provide insights into overcoming the growth barriers. Below, we discuss two disk substructures and how they may serve as favorable sites for dust growth.

\subsection{Planetesimal Formation in Dust Rings}\label{sec:1.1}

The Atacama Large Millimiter/submillimeter Array (ALMA) uncovered a diverse range of disk substructures in the (sub)millimeter dust continuum, among which rings and gaps are the most ubiquitous \citep{andrews+18,huang_etal18,andrews20}. These rings are commonly identified as likely venues for planetesimal formation \citep[e.g.,][]{Stammler_2019}, and can be formed and maintained by numerous mechanisms, such as planet-disk interactions \citep{gt79,lp86,lp93,Keppler+18,Haffert+19,pinte+23,bae+23}. To collect dust particles within rings, gas pressure bumps are typically required, as dust naturally migrates towards pressure maxima. These local pressure bumps have been seen in many numerical simulations with a planet carving one or multiple gaps \citep[e.g.,][]{pinilla+12,dong+15}. Furthermore, since dust particles within are expected to experience turbulent diffusion (\citealp[][hereafter D+18]{dullemond+18}, \citealp{rosotti+20}), dust rings have finite widths, as otherwise, particles would concentrate towards pressure maxima indefinitely.

The streaming instability (SI) \citep{yg05} is the leading mechanism for overcoming the bouncing, fragmentation, and radial drift barriers in planetesimal formation. The linear SI exhibits fastest growth rates when the relative drift velocity between the gas and dust induced by the background gas pressure gradient is in resonance with the phase velocity of an inertial wave  \citep{sh18b,sh18a,magnan+24}. The linear SI amplifies perturbations particularly fast for dust-to-gas ratios of unity or greater, and for marginally coupled particles, i.e. grains with stopping time $t_\mathrm{s}$ comparable to the local dynamical time scale $\Omega^{-1}$, or, in other words, a Stokes number $\mathrm{St} \equiv t_\mathrm{s}\Omega$ close to unity. Indeed, mm- or cm-sized solids that inhabit the top end of the expected dust size distribution in protoplanetary disks, possess a Stokes number of $\mathrm{St}\sim 0.1-1$. For such grains, the non-linear outcome of SI is characterized by strong, and rapidly emerging clumping of solids, within several local orbital timescales \citep[e.g.,][]{johansen_etal07,bs10a,carrera+15,yang+17,li+19,gerbig+20,schafer+24}. When particle clumps reach the Roche density, dust self-gravity enables the formation of planetesimals with $\sim 10-100$ km in size \citep[e.g.,][]{johansen+15,simon+16,schafer+17, klahr+20, GerbigLi2023}. 

% \textcolor{red}{ 
% A number of extensions have been applied to the SI. For instance, the externally driven turbulence and multiple dust species are likely of the most significance. The growth rate of the SI in the linear regime has been shown to reduce in the presence of external turbulence \citep{umurhan+20,cl20}. . When including multiple dust species, although linear theory has indicated that the fastest SI growth rate decreases with the number of dust species when dust-to-gas mass ratio $\epsilon<1$ \citep{krapp+19,zy21}, numerical simulations have shown that the maximum solid density converged with particle number as generically $\epsilon>1$ in the long term \citep{johansen+09,bs10a,schaffer+18,schaffer+21}.
% }

However, the picture of planetesimal formation in diffusive dust rings mediated by SI faces two problems. First, the presence of turbulent diffusion reduces the growth rate of the linear SI \citep{umurhan+20,cl20}, and may inhibit particle clumping during non-linear SI as seen in numerical simulations (with forced turbulence of $\alpha\sim10^{-3}$) \citep{lim+24}. Second, the pressure gradient vanishes at the center of a dust ring, removing the relative drift between gas and dust and depriving the SI of its energy source.  If the dust layer is subjected to appreciable levels of gas turbulence, it will nonetheless have a finite thickness, yet now without the ability to clump via drag instabilities, and is thus unable to form planetesimals. 

On the other hand, recent local non-ideal MHD shearing box simulations have demonstrated that particle clumping may be possible in pressure maxima \citep{xb22,xb22a}, if induced by MRI zonal flows that are commonly seen in MHD simulations \citep[e.g.,][]{johansen_etal09,bs14,rl19,cb21,cb22,hu+22}. Therefore, it is of importance to explore that, in the context of a pressure bump, 1) whether SI can still operate, and 2) whether new instabilities can set in to enhance local dust-to-gas ratio.

\subsection{Planetesimal Formation in Azimuthal Dust Clumps}\label{sec:}

Azimuthal crescent-shaped dust clumps have been observed in continuum of a handful of protoplanetary disks \citep[e.g.][]{vdmarel+13,huang_etal18,andrews20,vdmarel+21}. The crescent clumps are considered to be indicative of large, and potentially long-lived gas vortices. Similar to rings, these crescents provide ideal sites for overcoming the growth barriers by trapping particles now at the pressure maxima of vortices \citep{bs95}. The Rossby wave instability (RWI) is a promising mechanism to drive such large, lopsided vortices \citep{lovelace99}, as seen in numerical simulations of the RWI's non-linear evolution \citep{li+01}. 
The necessary condition of RWI is local extrema in the radial vortensity profile \citep{lovelace99,li_etal00,cy24}. Such local extrema can be readily produced by planets, due to a sharp transition in radial gas density at gap edges. Numerical simulations have indeed found the excitation of vortices in this context
\citep{zhu+14,zs14,hammer+17,huang+18,li+20,cimerman+23,ma+24}. 

The classic RWI is a purely gas-driven instability. While numerical investigations have shown that its nonlinear saturation, RWI effectively generates vortices and traps particles, its linear theory did not account for aerodynamic drag until very recently. \citet[][hereafter LB23]{lb23} is the first work to consider the effect of particles onto the RWI. As a pioneering study, they adopted a single-fluid formulation, where gas and dust were treated as one fluid \citep{ly17}. In a radially global shearing box, they employed a pressure bump background state mimicking the gas distribution of a ring, and discovered two distinct modes of the Dusty RWI (DRWI). Type I modes are similar to the gas RWI, in that growth rates peak for pure gas and decrease with increasing dust-to-gas mass ratio. Type II modes are novel and rely on the presence of dust. Moreover, \citetalias{lb23} performed shearing box simulations (on $H-$scales radially) to verify their linear theory and found that, while for Type I, the background dust ring structure is disrupted and vortices like gas RWI emerge, for Type II, the background ring persists and facilitate dust clumping. Thus, they concluded that Type II DRWI may be a new mechanism that can enhance local dust-to-gas ratios and promote planetesimal formation in protoplanetary disk rings seen by ALMA.

In the end of \S\ref{sec:1.1}, we discuss two directions that worth investigating at the location of ALMA rings (pressure bumps). The global linear analysis of the streaming instability in a pressure bump is explored in the first paper. In this study, we focus on the second direction by exploring the DRWI using a two-fluid formulation. 
The paper is organized as follows. In \S\ref{sec:me}, we introduce the basic equations and disk model. \S\ref{sec:nu} details the numerical method used. In \S\ref{sec:re}, we present the growth rates found for Type I and Type II DRWI modes. In \S\ref{sec:ALMA}, we apply parameters from DSHARP sources to the linear analysis and discuss the onset of DRWI in ALMA rings. Finally, we summarize the main findings in \S\ref{sec:c}.

% The dynamical processes within protoplanetary disks are exceedingly complex \citep[][]{armitage11}. Among the many factors shaping it, magnetic fields play a crucial role, influencing disk dynamics in several ways. Magnetized disk winds and laminar magnetic stress can efficiently extract angular momentum, driving accretion onto the protostars \citep[e.g.,][]{wbf21,cui+24debris}. The magneto-rotational instability \citep[MRI;][]{bh91}, weakened by non-ideal MHD effects, contributes to moderate levels of turbulence. Moreover, magnetically induced radial pressure variations can shape large-scale structure of the disk, potentially explaining observed annular substructures \citep[e.g.,][]{rl19,cb21,cb22} and azimuthal asymmetries \citep[e.g.,][]{cui+24,cw24,Hsu+24}.

% Despite the close connection of being dust grain traps, and the impact of magnetic fields on disk dynamics, linear theory has historically focused on modes of pure RWI \citep[e.g.,][]{lovelace99,ono_etal16}. It is only very recently that linear studies incorporating each of these two components have begun to emerge. Using a radially global model, \citet{cui+24,cw24} investigated the influence of poloidal and toroidal magnetic fields on RWI. They found that magnetism can either enhance or diminish RWI growth rates in the ideal MHD regime, whereas strong non-ideal MHD effects cause RWI growth rates to revert to hydrodynamic values. 

\section{Methods}\label{sec:me}

In this section, we present the basic equations and disk model. Specifically, they are dynamical equations \eqref{sec:de}, definition of parameters \eqref{sec:def}, disk model \eqref{sec:dm}, equilibrium solutions \eqref{sec:es}, and linearized equations \eqref{sec:le}. 

\subsection{Dynamical Equations}\label{sec:de}

We employ the unstratified shearing sheet approximation \citep{gl65,lp17}, which is a Cartesian representation of a small patch in the global disk. It is centered at $(r_0,\phi_0=\Omega t,z_0=0)$, and rotates around the central star with mass $M$ at Keplerian angular velocity $\Omega =\OmK(r_0)$. Cartesian coordinates $(x,y,z)$ denote radial, azimuthal, and vertical directions. Keplerian velocity appears as a linear shear in the shearing box,  $-q\Omega x \bb{e}_y$ with $q=3/2$. Moreover, we consider a two-fluid protoplanetary disk composed of gas and dust. The gas density, velocity, and pressure are denoted by $(\rg,\vg,P)$, and the pressureless \citep{lynch24}, single-species dust has density and velocity denoted by $(\rd,\vd)$. 
% In the following, we employ the velocity by subtracting the orbital shear, 
% \begin{equation}
% v'= v+\frac{3}{2} \Omega x \bb{e}_y, 
% \label{eq:1}
% \end{equation}
% and drop the prime on $v'$ to simplify the notation.

The dynamical equations of the gas-dust two-fluid framework are, 
\begin{equation}
\frac{\p\rg}{\p t} + \nabla\cdot\left(\rg\vg\right) = 0,
\label{eq:2}
\end{equation}
\begin{equation}
\frac{\p\rd}{\p t} + \nabla\cdot\left(\rd\vd\right) = 0,
%\nabla \cdot \bigg[D\rg\nabla\bigg(\frac{\rd}{\rg}\bigg)\bigg].
\label{eq:3}
\end{equation}
\begin{align}
\frac{\p\vg}{\p t} + \vg\cdot\nabla\vg 
= & - \frac{\nabla P}{\rg} - 2\bb{\Omega} \times\vg + 3\Omega ^2x\mathbf{e}_x + \frac{\epsilon}{\ts}(\vd-\vg) \nn \\
&  + \nu\nabla^2\vg + F(x)\mathbf{e}_y,
\label{eq:5}
\end{align}
\begin{align}
\frac{\p\vd}{\p t} + \vd\cdot\nabla\vd 
= & - 2\bb{\Omega} \times\vd + 3\Omega ^2x\mathbf{e}_x - \frac{1}{\ts}(\vd-\vg) \nn \\
& + \frac{1}{\rd}\nabla\cdot(\rd\bb{v}_\mathrm{dif}\bb{v}_\mathrm{dif}).
\label{eq:4}
\end{align}
In the above, $\epsilon=\rd/\rg$ is the dust-to-gas mass ratio, $\ts$ is the particle stopping time, $\nu$ is the kinematic viscosity, $\bb{v}_\mathrm{dif}$ is the dust diffusion velocity, and $F(x)$ is a forcing term in the $y$-direction. See the definitions of these parameters in \S\ref{sec:def} and \S\ref{sec:nds}.
The equation of state is set to be locally isothermal,
\begin{equation}
P=\rg\cs^2, 
\end{equation}
and $\cs$ denotes the isothermal sound speed. The units throughout are chosen to be $GM=\rho_0=r_0=1$. The gas pressure scale height is defined as $H \equiv \cs/\Omega $, with $H/r_0=0.05$.

\subsection{Definition of Parameters}\label{sec:def}

Dust dynamics is strongly affected by the coupling to the gas via the drag force in protoplanetary disks. For a large portion of the disk and particle sizes, the mean free path of the gas molecules is greater than the particle radius. This is called the Epstein drag law regime \citep{epstein24}. The particle stopping time, a timescale for the decay of the relative velocity between the dust and gas due to the drag force, in the Epstein regime is  
\begin{equation}
\ts = \frac{\rho_\mathrm{s}a}{\rg v_\mathrm{th}},
\label{eq:ts}
\end{equation}
where $\rho_\mathrm{s}$ is the material density of the particle, $a$ is the radius of the particle (treated as a hard sphere), and $v_\mathrm{th}=\sqrt{8/\pi}\cs$ is the gas mean thermal velocity \citep{birnstiel24}. Since we consider the motions of dust in the gas disk, the relevant timescale is the orbital period. Thereby, it is useful to define a dimensionless stopping time, or the so called Stokes number
\begin{equation}
\mathrm{St} = \ts\Omega .
\end{equation}
In this work, we focus on the regime where the dust and gas are well-coupled ($\mathrm{St\ll 1}$), in order to be consistent with the non-drift steady state solution adopted in \S\ref{sec:nds}.

Turbulent diffusion is a significant physical quantity in the context of annular substructures, as it determines the width of a ring \citepalias{dullemond+18}. Without it, dust particles will continue to migrate indefinitely towards the pressure maxima. In this work, gas turbulence is modeled by a viscous term in Equation \eqref{eq:5}, where $\nu=\alpha\cs H$ is the kinematic viscosity, and $\alpha$ is the Shakura-Sunyaev parameter \citep{ss73}. The dust diffusion term in Equation \eqref{eq:4} is modeled by the Reynolds-averaged mean-flow method, with a gradient diffusion approach to close the system \citep{binkert23}. 
The dust diffusion velocity is expressed by (\citealp{binkert23,tominaga+23}, \citetalias{lb23}),
\begin{equation}
\bb{v}_\mathrm{dif} = -D_\mathrm{d} \frac{\rg}{\rd}\nabla\frac{\rd}{\rg},
\end{equation}
where $D_\mathrm{d}$ is the particle diffusion coefficient. 
Note that in equations \eqref{eq:2}-\eqref{eq:4}, $\vd$ has two components: the time-averaged dust velocity and the dust diffusion velocity $\bb{v}_\mathrm{dif}$. Hence, the drag force terms includes the dust diffusion velocity.

It is common to relate the gas diffusion coefficient to particle diffusion coefficient \citep{yl07}. We treat dust and gas diffusion coefficients to be equal and constants, 
$D_\mathrm{d}=\nu=\mathrm{const}$.
Note that this approximation is only strictly valid when $\mathrm{St\ll 1}$, because $\nu/D_\mathrm{d}\sim 1+\mathrm{St^2}$ \citep{yl07}.

%The dust continuity equation takes the form of an advection-diffusion equation, where the second term on LHS is the advective term and on the RHS is a diffusive term \citep[e.g.,][]{mv84,cuzzi+93,Dubrulle+95,CL20}.  

\subsection{Disk Model}\label{sec:dm}

The disk model is established with a symmetric bump centered at $x=0$. The bump is specified by a Gaussian profile in the gas density and pressure,
\begin{equation}
\rho_{g0} = \rho_0\bigg[1+ A\exp\bigg(-\frac{x^2}{2w^2}\bigg)\bigg].
\label{eq:rg0}
\end{equation}
Here, $\rho_0$, $A$, $w$ denote background gas density, bump amplitude and width. The fiducial parameters are $A=1.5$ and $w=H$. Such a Gaussian bump can produce vortensity extrema in $x$, which is a necessary condition to excite the gas RWI \citep{cy24}. 

Background dust density can be obtained by setting up the disk equilibrium (\S\ref{sec:nds}), and is shown in Figure \ref{fig:rhod}. However, to ease the calculations, rather than directly using the dust density computed from equilibrium, we approximate the results with a Gaussian profile,  
\begin{equation}
\rho_{d0}* = \rho_0 \max\bigg[A_d \exp\bigg(-\frac{x^2}{2w_d^2}\bigg), \, 0.01\bigg],
\label{eq:rd0}
\end{equation}
where $A_d=\epsilon_\mathrm{max}(1+A)$, and $\epsilon_\mathrm{max}$ is the background dust-to-gas mass ratio at the bump center. The Gaussian profile is the analytical solution to the dust trapping problem \citepalias{dullemond+18}. For locations far from the bump center, we require $\epsilon=0.01$. This is done by setting a minimum $\rho_{d0}*=0.01$, as the minimum gas density $\rho_{g0}\sim 1$. 

\subsection{Equilibrium Solutions}\label{sec:es}

In the steady state, temporal derivatives vanish ($\p/\p t=0$), and axisymmetry is assumed ($\p/\p y=0$). Equations \eqref{eq:2}-\eqref{eq:4} then become
\begin{equation}
\frac{\p \vgx}{\p x} = -\frac{\p\ln\rg}{\p x}\vgx, 
\label{eq:A1}
\end{equation}
\begin{equation}
\frac{\p \vdx}{\p x} = -\frac{\p\ln\rd}{\p x}\vdx,
%+ \frac{1}{\rd}\frac{\p}{\p x}\bigg(D\rg\frac{\p \epsilon}{\p x}\bigg). 
\label{eq:A2}
\end{equation}
\begin{align}
\vgx\frac{\p \vgx}{\p x} = &-\frac{1}{\rg}\frac{\p P}{\p x} + 2\Omega \vgy + 3\Omega ^2x + \nu\frac{\p^2\vgx}{\p x^2} \nn \\
&+ \frac{\epsilon}{\ts}(\vdx-\vgx),
\label{eq:A3}
\end{align}
\begin{equation}
\vgx\frac{\p\vgy}{\p x} = - 2\Omega  \vgx + \mathrm{\nu}\frac{\p^2\vgy}{\p x^2} + \frac{\epsilon}{\ts}(\vdy-\vgy) + F,
\label{eq:A4}
\end{equation}
\begin{align}
\vdx\frac{\p\vdx}{\p x} = & +2\Omega \vdy + 3\Omega ^2x + \frac{1}{\rd}\frac{\p}{\p x}[\rd D_\mathrm{d}^2 (\p\ln \epsilon/\p x)^2] \nn \\
& - \frac{1}{\ts}(\vdx-\vgx),
\label{eq:A5}
\end{align}
\begin{equation}
\vdx\frac{\p\vdy}{\p x} = -2\Omega  \vdx- \frac{1}{\ts}(\vdy-\vgy),
\label{eq:A6}
\end{equation}
where we drop the subscription `0' for equilibrium quantities to simplify the notation.

%\subsubsection{Drift Solution}\label{sec:ds}

In the subsection below, we derive the non-drift steady state solution, that is different from the drift solution adopted in \citet{yg05} for SI. By ignoring viscosity and forcing ($\nu=F=D_\mathrm{d}=0$), eqs. \eqref{eq:A1}-\eqref{eq:A6} resemble the steady state equations of SI \citep{yg05}. The solution of it was adopted from \citet{NSH86}, which contains a velocity drift between gas and dust, with relative radial velocity $\sim O(\mathrm{St}f)$ and relative azimuthal velocity $\sim O(\mathrm{St}^2f^2)$, when $\mathrm{St}^2f^2\ll 1$, where $f=\rg/(\rg+\rd)$. 
While the onset of SI relies on the relative drift between gas and dust, this is not required for the (D)RWI, because the classic RWI only involves the gas component \citep{lovelace99}. In fact, our non-drift solution effectively filters out the SI. 
%We also note that similar to the drift solution used in \citet{yg05}, the non-drift solution only satisfies the momentum equations \eqref{eq:A3}-\eqref{eq:A6}, but not the continuity equations \eqref{eq:A1}-\eqref{eq:A2}.   

\subsubsection{Non-Drift Solution}\label{sec:nds}

We present a simple solution to equations \eqref{eq:A1}-\eqref{eq:A6} without velocity drift.
This non-drift solution reads
\begin{equation}
v_\mathrm{gx0}=v_\mathrm{dx0}=0,  \qquad v_\mathrm{gy0}=v_\mathrm{dy0}.
\end{equation}
Physically, it corresponds to a regime where the Stokes number is sufficiently small. As we will discuss in \S\ref{sec:re}, the two types of DRWI modes exhibit more intriguing behaviors when $\mathrm{St \leqslant 1}$. For $\mathrm{St = 10}$, the growth rates closely resemble those of the classic gas RWI. However, it is important to note that, strictly speaking, our results are only valid in the regime $\mathrm{St}f\ll 1$.
To further simplify the solution, we impose the condition that both radial velocities vanish in gas and dust. Consequently, all the left hand sides of momentum equations become zero. Now, equations \eqref{eq:A1}-\eqref{eq:A6} are simplified to three relations, where $v_\mathrm{gy0}, v_\mathrm{dy0}, F, \rho_\mathrm{d0}$ can be computed. 

First, the steady state $y$-velocities can be directly found by Equation \eqref{eq:A3},
\begin{equation}
v_\mathrm{gy0} = v_\mathrm{dy0} = -\frac{3}{2} \Omega  x + \frac{1}{2\Omega }\frac{1}{\rho_\mathrm{g0}}\frac{\p P_0}{\p x}.
\label{eq:vy}
\end{equation}
This is simply the standard Keplerian profile with a pressure gradient on top of it, commonly seen in the pure gas system. In a pressure bump, the $y$-velocity is sub-Keplerian for $x>0$ and super-Keplerian for $x<0$. Second, the forcing $F$ is used to balance the turbulent diffusion due to the shear of $\vgy$ in $x$. By Equations \eqref{eq:A4} and \eqref{eq:vy}, $F$ is given by
\begin{equation}
F(x) = -\nu\frac{\p^2v_\mathrm{gy0}}{\p x^2} = -\nu\frac{c_s^2}{2\Omega }\frac{\p^3 \ln\rho_\mathrm{g0}}{\p x^3}.
\label{eq:F}
\end{equation}
Since $F(x)$ is a forcing term, it only appears in the steady states, but not involved in the linearized equations. 

\begin{figure}
    \centering
    \includegraphics[width=0.5\textwidth]{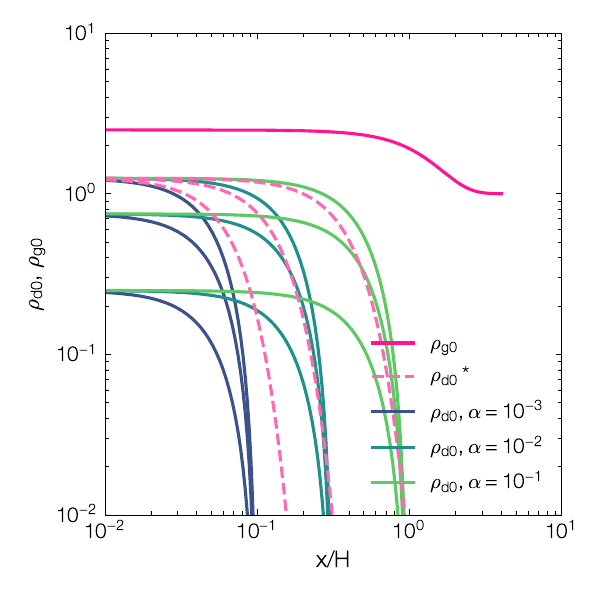}
    \caption{Gas and dust density distributions in log-log scale. Gas density ($\rho_\mathrm{g0}$) is plotted with Equation \eqref{eq:rg0} (solid pink). Dust density ($\rho_\mathrm{d0}$) is plotted by solving Equation \eqref{eq:rhod} (green and blue), for different turbulence levels $\alpha\in[10^{-3},10^{-2},10^{-1}]$ (left to right) and dust-to-gas mass ratios $\epsilon_\mathrm{max}\in[0.5,0.3,0.1]$ (high to low). The dashed lines show $\rho_{d0}*$ defined in Equation \eqref{eq:rd0} with $w_\mathrm{d}/H\in[0.05, 0.1, 0.3]$ at $\epsilon_\mathrm{max}=0.5$ (left to right).
    }
    \label{fig:rhod}    
\end{figure}

Lastly, dust density profile $\rd$ can be computed by comparing the dust diffusion term in Equation \eqref{eq:A5} and the gas pressure gradient term in Equation \eqref{eq:A3}. No relative drift between $\vgy$ and $\vdy$ indicates that these two terms should exactly equal,
\begin{equation}
\frac{1}{\rho_\mathrm{d0}}\frac{\p}{\p x}[\rho_\mathrm{d0} D_\mathrm{d}^2 (\p\ln \epsilon_0/\p x)^2] +\frac{1}{\rho_\mathrm{g0}}\frac{\p P_0}{\p x}=0.
\label{eq:rhod}   
\end{equation}
Dust density profile is computed by solving the above ordinary differential equation (ODE) as an initial value problem (IVP), where the initial condition is given at the center of the box (slightly shifted from $x=0$ to avoid singularity), and the IVP is solved for $x>0$. The detailed numerical method of solving this ODE can be found in Appendix \ref{app:rhod}. 

Figure \ref{fig:rhod} shows the gas and dust density distributions. The gas density ($\rho_\mathrm{g0}$) is directly plotted from Equation \eqref{eq:rg0}. The dust density ($\rho_\mathrm{d0}$) is obtained by solving Equation \eqref{eq:rhod}, for different levels of turbulence $\alpha\in[10^{-3},10^{-2},10^{-1}]$ from left to right \citep{rosotti+23}, and dust-to-gas mass ratios $\epsilon_\mathrm{max}\in[0.5,0.3,0.1]$ from high to low. Note that our $\rho_\mathrm{d0}$ does not depend on the Stokes number, because there is no drift in equilibrium. Dashed pink curves denote $w_\mathrm{d}/H\in[0.05, 0.1, 0.3]$ from left to right for $\rho_\mathrm{d0}*$ defined in Equation \eqref{eq:rd0}. For simplicity, we utilize $\rho_\mathrm{d0}*$ rather than $\rho_\mathrm{d0}$ computed from IVP as the equilibrium dust density when conducting the linear analysis throughout this work.
Our tests show that the Gaussian dust density profile (dashed curves) does not qualitatively alter the results compared to the numerically computed profile (solid green curves).

\subsection{Linearized Equations}\label{sec:le}

Consider small perturbations to eqs. \eqref{eq:2}-\eqref{eq:4}, such that for example,  $\bb{v} = \bb{v}_0+\delta\bb{v}(x,y,t)$, $\rho = \rho_0+\delta\rho(x,y,t)$. We linearize these equations by considering Eulerian perturbations $\propto f(x)\exp(i\ky y-i\omega t)$, where $\ky$ is the $y$-wavenumber, and $\omega=\omega_r+i\gamma$ is the mode frequency, where $\gamma$ denotes the growth rate. 
We drop the subscription `0' to simplify the notation for the linearized equations.
The linearized continuity, momentum, and energy equations are 
\begin{align}
-i\Delta\omega\delta\rg & + \frac{\p [\rg\delta\vgx]}{\p x} + i\ky\rg\delta\vgy=0,
\end{align}

\begin{align}
-i\Delta\omega_\mathrm{d}\delta\rd & + \frac{\p [\rd\delta\vdx]}{\p x} + i\ky\rd\delta\vdy=0,
\end{align}

\begin{align}
-i\Delta\omega\delta\vgx & - 2\Omega \delta\vgy + \frac{1}{\rg}\frac{\p\delta P}{\p x} - \frac{\delta\rg}{\rg^2}\frac{\p P}{\p x} \nn \\
& - \nu\bigg(\frac{\p^2}{\p x^2} -\ky^2\bigg)\delta\vgx - \frac{1}{\ts}\frac{\rd}{\rg} (\delta\vdx-\delta\vgx) \nn \\
& = 0,
\end{align}

\begin{align}
-i\Delta\omega\delta\vgy & + \delta\vgx\frac{\p\vgy}{\p x} + 2\Omega \delta\vgx + \frac{i\ky}{\rg}\delta P  \nn \\
& - \nu\bigg(\frac{\p^2}{\p x^2} -\ky^2\bigg)\delta\vgy - \frac{1}{\ts}\frac{\rd}{\rg}(\delta\vdy-\delta\vgy) \nn \\ 
& = 0,
\end{align}

\begin{align}
-i\Delta\omega_\mathrm{d}\delta\vdx - 2\Omega \delta\vdy + \frac{1}{\ts}(\delta\vdx-\delta\vgx) - D_\mathrm{dx} = 0,
\label{eq:26}   
\end{align}

\begin{align}
-i\Delta\omega_\mathrm{d}\delta\vdy & + \delta\vdx\frac{\p\vdy}{\p x} + 2\Omega \delta\vdx + \frac{1}{\ts}(\delta\vdy-\delta\vgy) \nn \\
& - D_\mathrm{dy} = 0,
\label{eq:27}   
\end{align}
\begin{align}
\delta P = \delta\rg \cs^2,
\end{align}
where $\Delta\omega=\omega-\ky\vgy$ and $\Delta\omega_\mathrm{d}=\omega-\ky\vdy$. 
%The dust diffusion terms $D_\mathrm{dx},D_\mathrm{dy}$ are expressed in Appendix \ref{app:le}.

The linearized dust diffusion terms in equations \eqref{eq:26} and \eqref{eq:27} are 
\begin{align}
D_\mathrm{dx} = & + \frac{1}{\rd}\frac{\p}{\p x}(2\rd v_\mathrm{diffx} v_\mathrm{diffx}'+\delta\rd v_\mathrm{diffx}^2)  \nn \\
& - \frac{\delta\rd}{\rd^2}\frac{\p}{\p x}(\rd v_\mathrm{diffx}^2) + i\ky v_\mathrm{diffx} v_\mathrm{diffy}',
\end{align}

\begin{align}
D_\mathrm{dy} = \frac{1}{\rd}\frac{\p}{\p x}(\rd v_\mathrm{diffx}v_\mathrm{diffy}'),
\end{align}
where the steady state and perturbed dust diffusion velocities are 

\begin{align}
v_\mathrm{diffx} = D_\mathrm{d}\frac{\rg}{\rd}\frac{\p}{\p x}\frac{\rd}{\rg},
\end{align}

\begin{align}
v_\mathrm{diffy} = 0,
\end{align}

\begin{align}
v_\mathrm{diffx}' = D_\mathrm{d}\bigg[\frac{\rg}{\rd}\frac{\p}{\p x}\bigg(\frac{\delta\rd}{\rg} - \frac{\rd\delta\rg}{\rg^2}\bigg) + \bigg(\frac{\delta\rg}{\rd} - \frac{\rg\delta\rd}{\rd^2}\bigg) \frac{\p}{\p x}\frac{\rd}{\rg}\bigg] ,
\end{align}

\begin{align}
v_\mathrm{diffy}' = D_\mathrm{d}i\ky\bigg(\frac{\delta\rd}{\rd} - \frac{\delta\rg}{\rg}\bigg).
\end{align}

\section{Numerics}\label{sec:nu}

We solve linearized equations numerically via the pseudospectral method \citep[see e.g.,][]{cl21,cui+24}, also known as the orthogonal collocation method. This technique approximates solutions to differential equations by a weighted sum of orthogonal basis functions, typically trigonometric functions or orthogonal polynomials, up to a certain degree \citep{boyd}. 
The linearized differential equations presented in \S\ref{sec:le} formulate the standard linear eigenvalue problem (EVP). In a generalized matrix form, it is written as
\begin{equation}
A\vec{x} = L\vec{x} + \omega M\vec{x} = 0,
\end{equation}
where $\omega$ is the eigenvalue, and $\vec{x} = [\vec{\delta\vgx},\vec{\delta\vgy},\vec{\delta \vdx},\vec{\delta\vdy},\vec{\delta\rg},\vec{\delta\rd}]^\mathrm{T}$ is a vector of eigenfunctions with six perturbed quantities. Here, $A$, $L$, $M$ are $6N\times 6N$ sized matrices, with $L$ consists of linear operators. 

We employ \textsc{Dedalus}\footnote{\url{https://dedalus-project.org/}} a general purpose spectral code to solve the EVP \citep{burns+20}. Fourier series are chosen as the orthogonal basis functions to expand the eigenfunctions, which naturally enforce periodic boundary conditions (a comparison of boundary conditions between this work and previous studies is provided in Appendix \ref{app:bc}). The computational domain spans $x\in[-4H,4H]$ and is discretized into $N$ points with uniform spacing. We employ the dense solver method via \texttt{solve\_dense}, which converts matrix $A$ into dense arrays, and utilize \texttt{scipy.linalg.eig} routine from Python, that can solve the EVP directly. A numerical resolution of $N=251$ is adopted. The resolution is increased to $N=401$ to ensure convergence and accuracy of the results when necessary.

\section{Results}\label{sec:re}

\subsection{Classic RWI}\label{sec:grwi}

\begin{figure}
    \centering
    \includegraphics[width=0.5\textwidth]{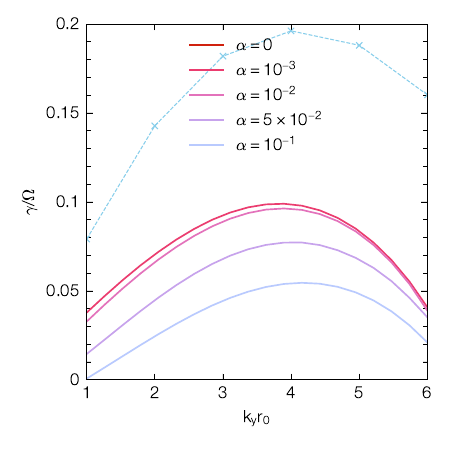}
    \caption{Solid curves: growth rates of the pure gas RWI for different viscous $\alpha$ values. Dashed curve: comparision with HGB model of Figure 10 in \citet{li_etal00}.
    }
    \label{fig:alp}    
\end{figure}

We begin by computing the linear growth rates of the pure gas RWI, using only the gas components of equations \eqref{eq:A1}–\eqref{eq:A6} within the compressible shearing sheet framework adopted in this work. Our goal is to reproduce previously reported growth rates for the gas-only RWI to strengthen the reliability of our results. However, no existing study employs exactly the same model setup. For example, \citetalias{lb23} used a shearing sheet, but their formulation combined gas and dust into a single fluid, complicating direct comparison.
Instead, we compare our results with those of \citet{li_etal00}, who studied the RWI in cylindrical coordinates using an adiabatic equation of state. Specifically, we refer to their homentropic Gaussian bump (HGB) model shown in Figure 10. Using the same parameters, we find slightly lower growth rates, for instance, $\gamma/\Omega  = 0.196$ at $m=4$ and $\gamma/\Omega  = 0.188$ at $m=5$, but the overall trend is consistent with theirs (dashed curve, Figure \ref{fig:alp}).

Then, we compute the growth rates of gas RWI for fiducial parameters chosen in this work, $A=1.5$ and $w=H$. 
Figure \ref{fig:alp} shows the growth rates for different viscosity strengths. In the absence of viscosity ($\alpha=0$), it is found that the maximum growth rate peaks at $k_{y,\mathrm{max}}r_0=3.89$, and $\gamma_\mathrm{max}/\Omega =0.0989$. The oscillation frequencies $\omega_r$ for all modes approach zero, but note that $\omega_r\neq 0$, which means the corotation radius is close to $x=0$. With the presence of viscosity ($\alpha\neq 0$), the growth rates decrease monotonically for all $\ky$. However, this decreasing only becomes pronounced for sufficiently strong viscosity ($\alpha\gtrsim 10^{-2}$). For $\alpha=10^{-1}$, the wavenumber of maximum growth rate shifts to higher $k_{y,\mathrm{max}}r_0=4.16$, and the growth rate drops to $\gamma_\mathrm{max}/\Omega =0.0545$.

\subsection{Type I DRWI}\label{sec:drwiI}

\begin{figure*}
    \centering
    \includegraphics[width=\textwidth]{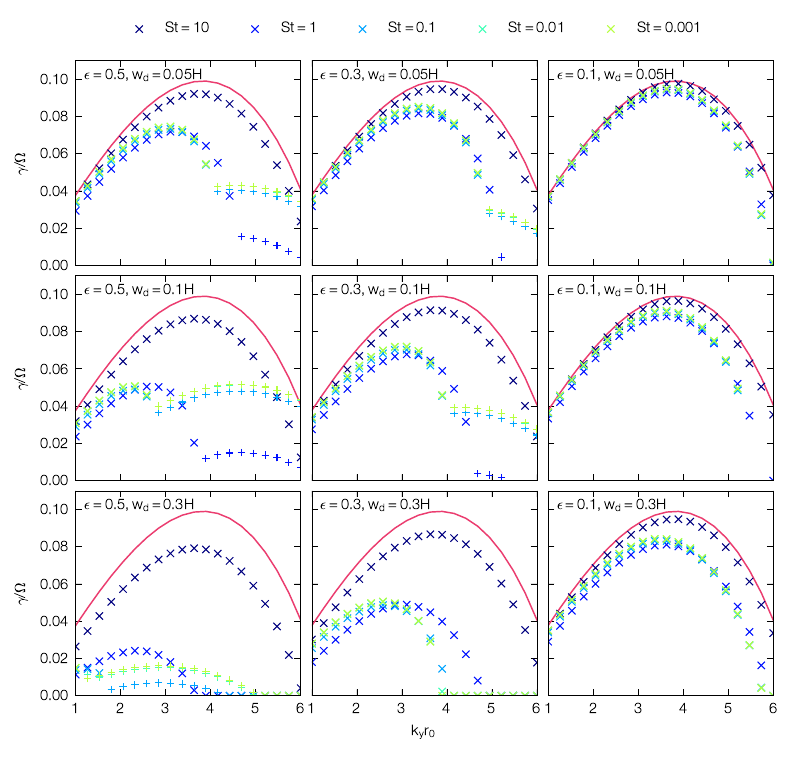}
    \caption{DRWI Growth rates for $\alpha=10^{-3}$ of different dust-to-gas density ratios at the bump center $\epsilon_\mathrm{max}=A_d/A\in[0.5,0.3,0.1]$ (from left to right), and different dust bump widths $w_d/H\in[0.05, 0.1, 0.3]$ (from top to bottom).
    Type I DRWI modes are denoted by crosses, and type II DRWI modes are denoted by plus symbols.
    Red curves denote pure gas RWI growth rates (\S\ref{sec:grwi}).
    }
    \label{fig:alp1e-3}    
\end{figure*}

\begin{figure}
    \centering
    \includegraphics[width=0.5\textwidth]{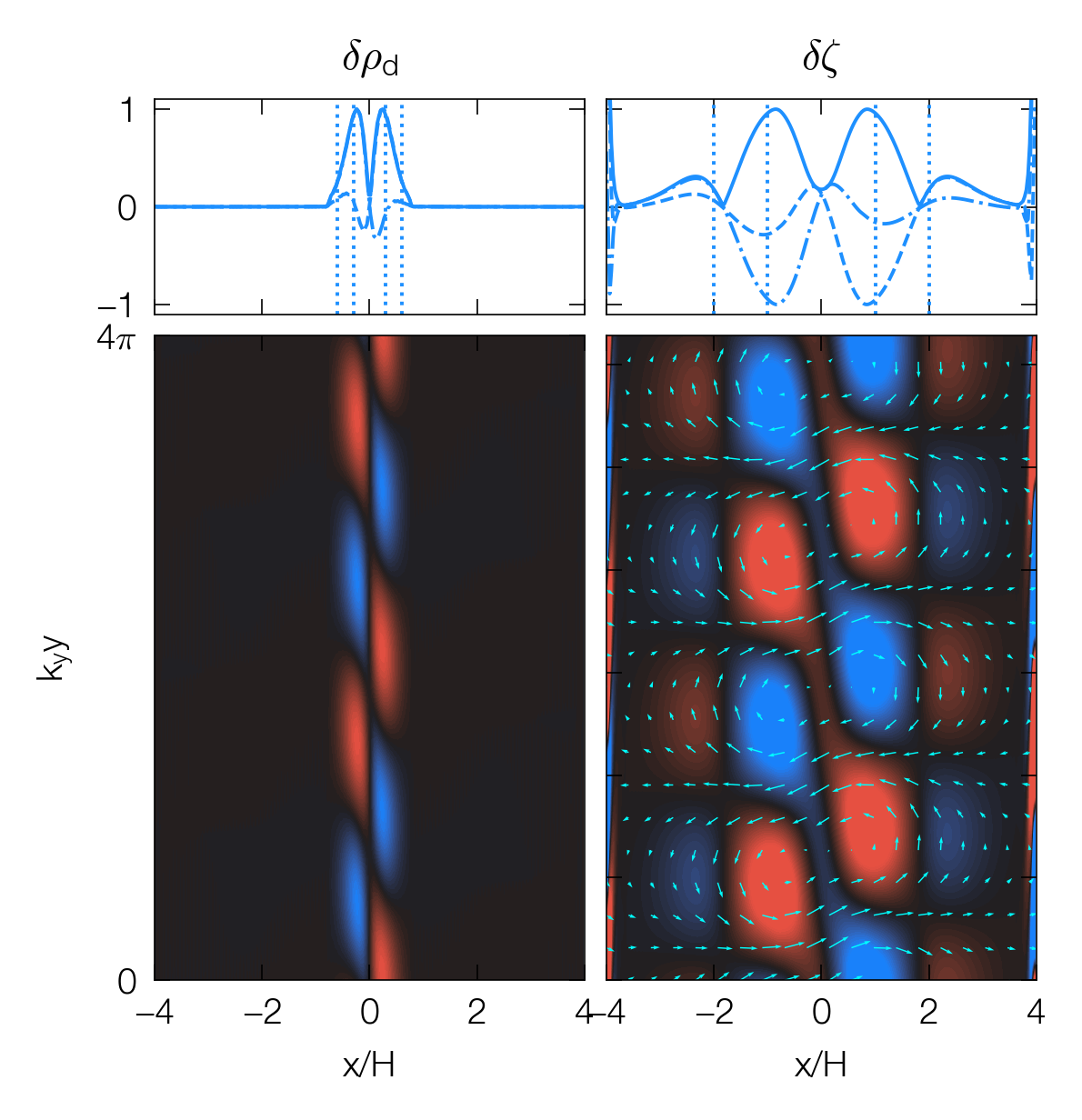}
    \caption{Eigenfunctions of Type I DRWI modes. 
    First row: normalized real (dashed), imaginary (dash-dotted), and absolute (solid) part of the perturbed quantities.
    Dashed vertical lines denote the locations of $x=\pm w_d, \pm 2w_d$ (left column) or $x=\pm w, \pm 2w$ (right column). 
    Second row: amplitude of perturbed quantities in the $(x,y)-$plane.
    Red (blue) denotes positive (negative) values. Arrows denote the perturbed velocity field.
    Parameters: $A=1.5$, $w=H$, $\alpha=10^{-3}$, $k_yr_0=3$, $\epsilon_\mathrm{max}=0.1$, $w_d/H=0.3$, $\mathrm{St=0.1}$, and $\omega/\Omega =1.3\times 10^{-5}+0.081i$. 
    }
    \label{fig:ef}    
\end{figure}

\begin{figure*}
    \centering
    \includegraphics[width=\textwidth]{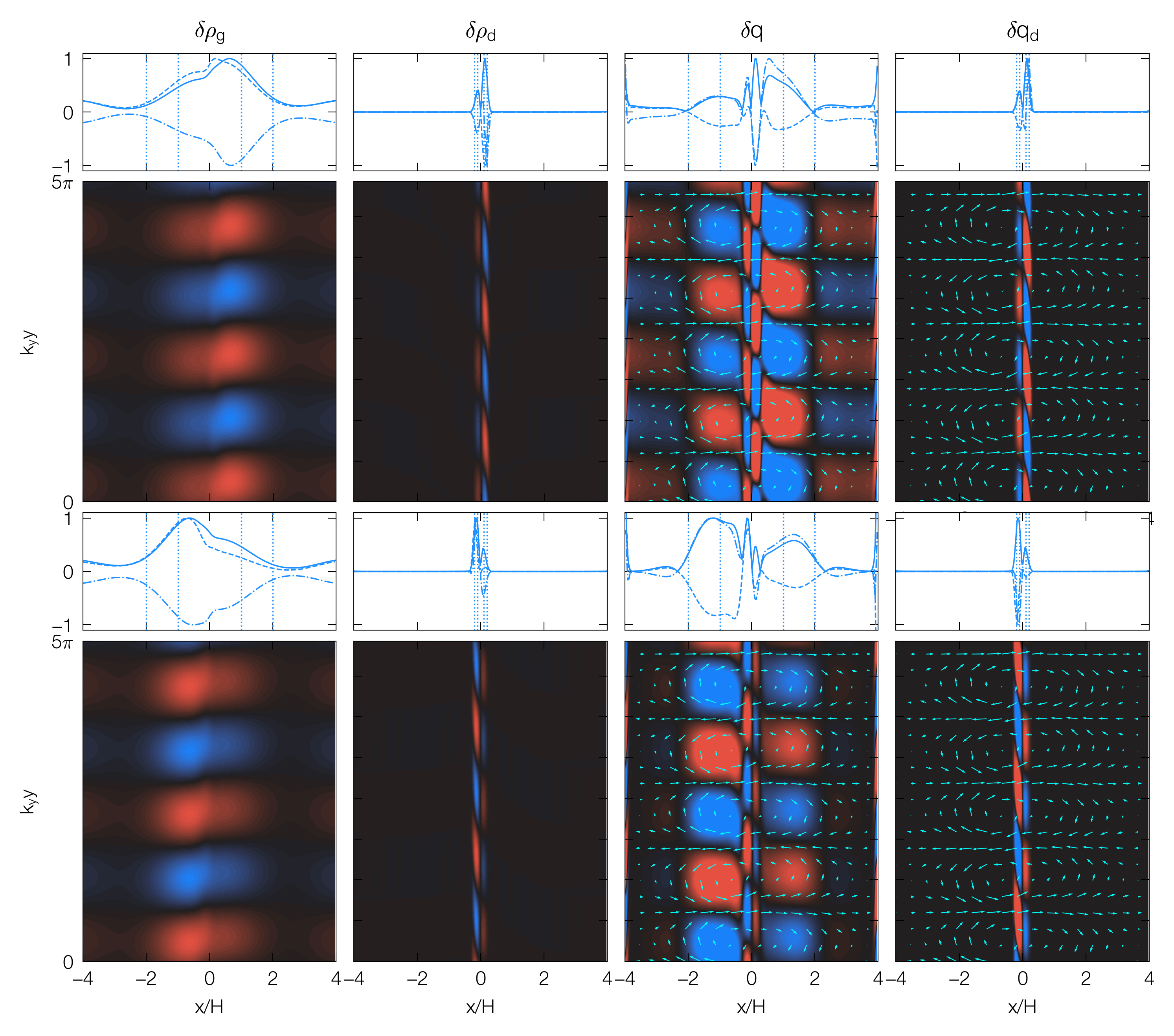}
    \caption{Eigenfunctions of a pair of Type II DRWI modes. 
    Top rows (right mode): $A=1.5$, $w=H$, $\alpha=10^{-3}$, $k_yr_0=5$, $\epsilon_\mathrm{max}=0.5$, $w_d/H=0.1$, $\mathrm{St=0.01}$, and $\omega/\Omega =-0.0955+0.047i$. Bottom rows (left mode): $\omega/\Omega =0.0955+0.047i$.
    }
    \label{fig:efII}    
\end{figure*}

\begin{figure}
    \centering
    \includegraphics[width=0.5\textwidth]{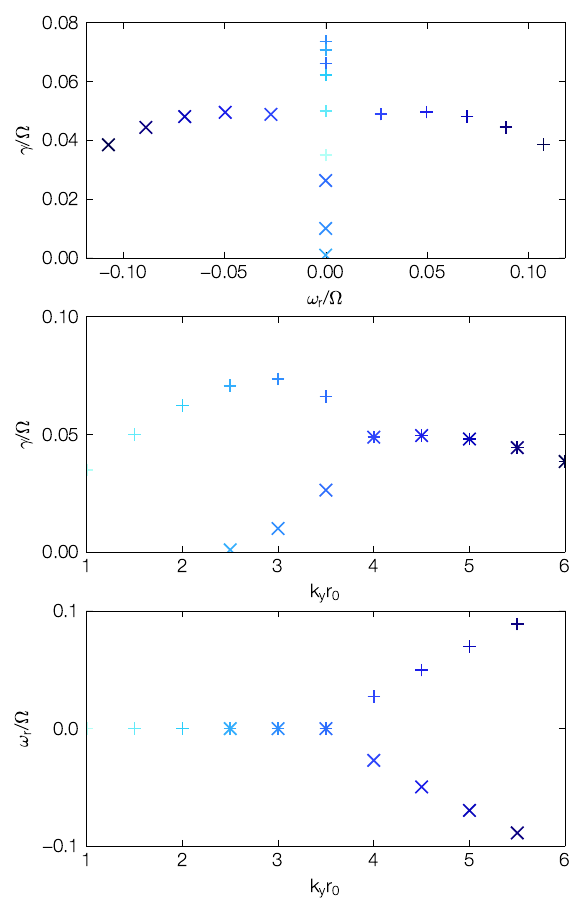}
    \caption{Growth rates $\gamma$ and oscillation frequencies $\omega_r$ for different wavenumbers $\ky$. Parameters: $A=1.57$, $w=H$, $\alpha=10^{-3}$, $\epsilon_\mathrm{max}=0.3$, $w_d=0.1$, and $\mathrm{St=0.01}$. 
    }
    \label{fig:sw}    
\end{figure}

Figure \ref{fig:alp1e-3} shows the DRWI growth rates with the presence of dust. The fiducial parameters are $A=1.5$, $w=H$ and viscosity $\alpha=10^{-3}$. We conduct parameter survey on the dust-to-gas density ratios at bump center $\epsilon_\mathrm{max}=A_d/A\in[0.1,0.3,0.5]$, widths of the dust bump $w_d/H\in[0.05,0.1,0.3]$\footnote{Note that for $\alpha=10^{-3}$, $w_d/H=0.05$ more closely matches the numerically computed dust equilibrium profile (see Figure \ref{fig:rhod}).}, and the Stokes number $\mathrm{St}\in[0.001,0.01,0.1,1,10]$.
Two distinct DRWI modes are found (Type I and Type II). Type I modes are dust modified RWI shown as crosses in Figure \ref{fig:alp1e-3}.

For very weak coupling between dust and gas ($\mathrm{St=10}$), the growth rates of Type I modes almost resemble the pure gas RWI results, shown in all panels of Figure \ref{fig:alp1e-3}. This is especially clearly seen in the right column, where dust-to-gas mass ratio, $\epsilon_\mathrm{max}=0.1$, is low in a dust ring. 
For $\mathrm{St}=1$, the growth rates decrease compared to $\mathrm{St}=10$. However, for smaller Stokes numbers $\mathrm{St}<1$, the growth rates do not drop significantly. 
Moreover, higher dust-to-gas ratios $\epsilon_\mathrm{max}$ and wider dust bumps $w_d$ tend to diminish the growth of Type I modes. 
%This finding is explained by the instability mechanism, when analyzing vortensity budget in \S\ref{sec:im}.

In Figure \ref{fig:alp1e-3}, Type I DRWI modes are present across a broad parameter space of $\epsilon_\mathrm{max}$, $w_d$, and St, emerging in all panels. For large dust-to-gas ratios $\epsilon_\mathrm{max} = 0.3,0.5$ and narrow dust rings $w_d/H=0.05,0.1$, Type I modes are confined to relatively small azimuthal wavenumbers ($k_yr_0 \leq 5$). This is because that at larger $k_y$, Type I modes are suppressed, and the Type II modes set in. Notably, Type I and Type II modes never coexist; for a given $\ky$, only one mode type can emerge, contrary to the findings in \citetalias{lb23} (see \S\ref{sec:comp}).

Figure \ref{fig:ef} shows the eigenfunctions of Type I modes for $\epsilon_\mathrm{max}=0.1$, $w_d/H=0.3$, and $\mathrm{St=0.1}$ (lower right column of Figure \ref{fig:alp1e-3}). We present the perturbed dust density $\delta\rd$, and perturbed vorticity $\delta\zeta$ (defined in Equation \eqref{eq:pz}), respectively.
Dashed vertical lines denote the locations of $x=\pm w_d, \pm 2w_d$ (left column) or $x=\pm w, \pm 2w$ (right column). It is evident that the perturbed vorticity, along with the perturbed gas density (not shown), are confined within the background gas bump, and the perturbed dust density remains within the background dust bump. The perturbed vorticity behaves similarly to the pure gas RWI, exhibiting a pair of counter-propagating Rossby waves across the corotation (see Figure 16 of \citet{ono_etal16} and \S3.3 of \citet{cy24}).

\subsection{Type II DRWI}\label{sec:drwiII}

Type II DRWI modes are shown as plus symbols in Figure \ref{fig:alp1e-3}. They always appear in pairs. For each pair of modes, they share the same growth rate $\gamma$, but possess oppositely-signed oscillation frequencies $w_r$. 
Type II modes emerge when dust and gas are well coupled. For $\mathrm{St=10}$, no Type II modes are found in Figure \ref{fig:alp1e-3}. When $\mathrm{St<1}$, the growth rates are generically larger than $\mathrm{St=1}$ modes. In addition, Type II modes are excited when dust content is high, $\epsilon_\mathrm{max}=0.3,0.5$, or the width of the dust bump is narrow, $w_d/H=0.05,0.1$.
Type II modes set in for relatively large azimuthal wavenumbers $\ky$ compared to Type I modes. 

Figure \ref{fig:efII} shows the eigenfunctions of a pair of Type II modes for $\epsilon_\mathrm{max}=0.5$, $w_d=0.1$, and $\mathrm{St=0.1}$ (middle left panel of Figure \ref{fig:alp1e-3}). We present the perturbed gas density $\delta\rg$, dust density $\delta\rd$, gas vortensity $\delta q$, and dust vortensity  $\delta q_\mathrm{d}$, respectively (see definitions of perturbed vortensities in \S\ref{sec:im}). 
Again, the perturbed gas density is clearly confined by the background gas bump, while the perturbed dust density is confined by the background dust bump. However, unlike Type I modes, the perturbed gas vortensity in this case exhibits a pair of thin waves across $x = 0$, situated between the Rossby waves. The presence of these narrow features serves as an additional signature of Type II modes, complementing their characteristic oppositely signed frequencies.

\subsection{Transition of DRWI modes}\label{sec:sm}

In Figure \ref{fig:alp1e-3}, we find that at a given $\ky$, Type I and Type II DRWI modes never coexist. Only one type of mode can be excited for a given $\ky$. Here, we demonstrate how the two types of modes transition from one to the other. We consider Type I modes as those with corotation radius close to the bump center ($x=0$), and Type II modes as those two modes share the same growth rate but have oppositely-signed oscillation frequencies.
Figure \ref{fig:sw} shows the growth rates $\gamma$ and oscillation frequencies $\omega_r$, for different wavenumbers $\ky$ denoted by different colors. For $\ky r_0\leq 2$, only one Type I mode is present (plus symbols), with $w_r/\Omega $ approaching zero. For $2<\ky r_0<4$, a new Type I mode sets in (crosses). The growth rate of this mode starts from zero and gradually increases with $\ky$, and the oscillation frequency is different from the first Type I mode, though still close to zero. When two Type I modes show up, we select the one with the higher growth rate for our analysis. At $\ky r_0=4$, the two different Type I modes transition to a pair of Type II mode. The corotation radii of Type II modes become far from the bump center with increasing $\ky$.  

\subsection{Instability Mechanism}\label{sec:im}

The pure gas RWI mechanism is most clearly reflected in the perturbed vortensity $\delta q$. A pair of counter-propagating Rossby waves forms across corotation \citep{Heifetz99}. This wave pair is shifted in the $y$-direction, and remains radially symmetric about corotation, consistent with the symmetric potential of the Gaussian bump shown in Figure 1 of \citet{cy24}. The $y$-direction phase shift causes gas flow passing through the background vorticity minimum ($x = 0$) to primarily enter regions of negative vorticity perturbation, and vorticity maximum to primarily enter regions of positive vorticity perturbation, which is a runaway process \citep{ono_etal16,cy24}. 

The above mechanism can be used to explain the Type I DRWI modes shown in Figure \ref{fig:ef}. A pair of counter-propagating Rossby waves is located on each side of the corotation. The amplitude of $\delta\zeta$ is symmetric about $x$, and the sign is consistent with the arrows and the background vorticity. In contrast, the Type II modes shown in Figure \ref{fig:efII} cannot be explained by this mechanism. Specifically, the sign of $\delta q$ and the direction of the arrows are inconsistent for the thin waves, suggesting that advection alone does not dominate the driving mechanism of Type II modes.

To clarify the physical mechanism of the Type I and Type II modes, we analyze the source of the perturbed gas and dust vortensities. We derive the vortensity equation in Appendix \ref{app:vor}. The gas vorticity $\zeta$ and vortensity $q$ in our shearing sheet model are defined as
\begin{equation}
\zeta = (\nabla\times \vg)_z,
\end{equation}
\begin{equation}
q = (2\Omega  +\zeta)/\rg,
\end{equation}
and the corresponding perturbed quantities are
\begin{equation}
\delta\zeta=(\nabla\times\delta\vg)_z,
\label{eq:pz} 
\end{equation}
\begin{equation}
\delta q = \delta\zeta/\rho_\mathrm{g0} - (2\Omega +\zeta)\delta\rg/\rho_\mathrm{g0}^2.
\end{equation}
Appendix \ref{app:vor} gives the vortensity equation,
\begin{equation}
\frac{Dq}{Dt} = \frac{S}{\rg},
\end{equation}
and the perturbed vortensity equation 
\begin{equation}
(-i\omega + i\ky v_\mathrm{gy0})\delta q + \delta\vgx \frac{\p q_0}{\p x} \approx S'_\mathrm{drag},
\label{eq:pv} 
\end{equation}
where  
\begin{equation}
S'_\mathrm{drag} = \frac{1}{\rg}\frac{\epsilon}{\ts}\nabla\times(\delta\vd-
\delta\vg) .
\end{equation}

\begin{figure}
    \centering
    \includegraphics[width=0.5\textwidth]{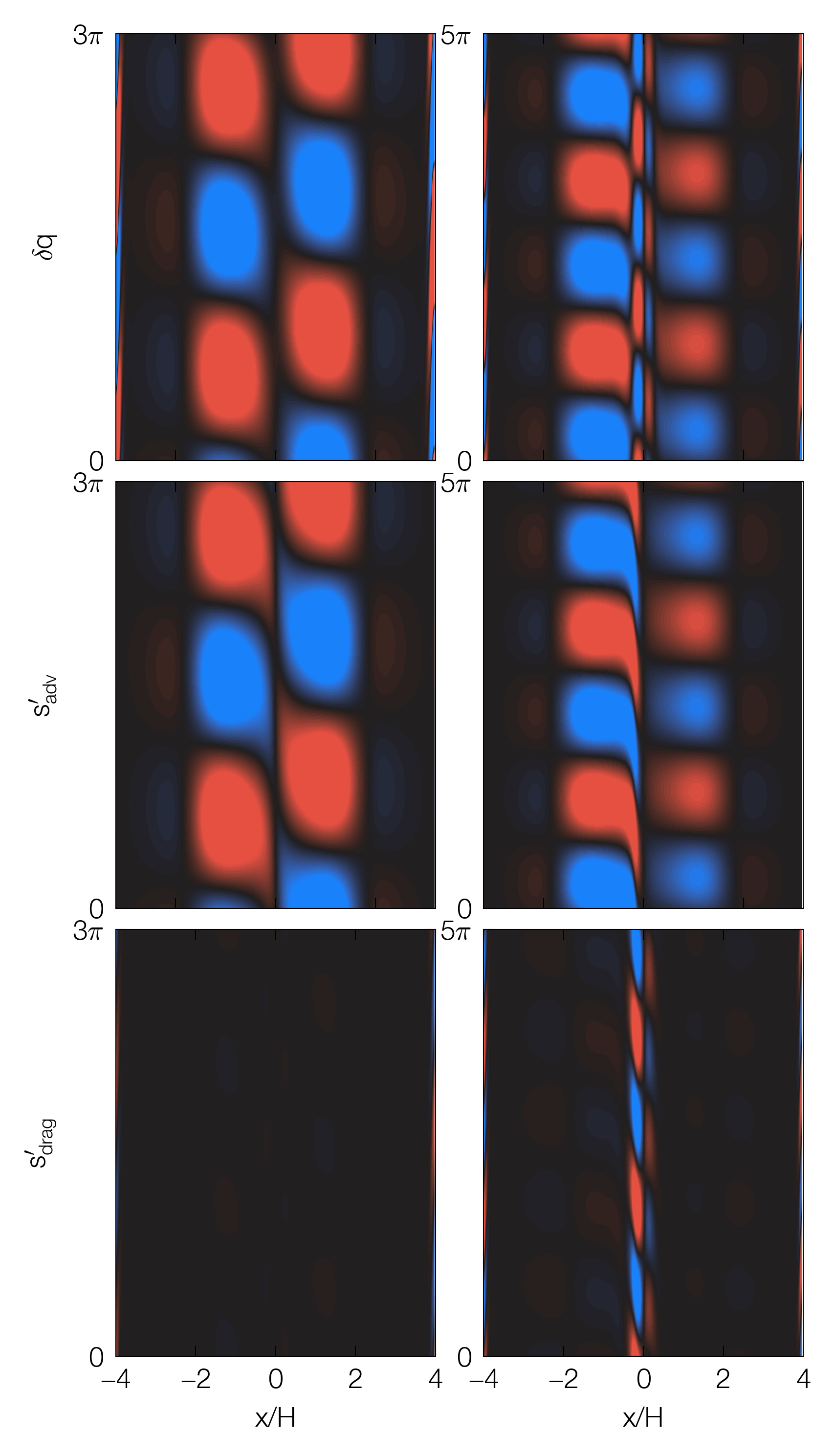}
    \caption{
    Terms in the perturbed vortensity equation (\ref{eq:pv}) shown in $(x/H-\ky y)$ plane. 
    Left column: Type I mode in Figure \ref{fig:ef}.
    Right column: Type II mode in Figure \ref{fig:efII} (left mode).
    Top to bottom row: perturbed vortensity $\delta q$, advection term, and drag force term. 
    All panels in each column have the same colorbar range.
    }
    \label{fig:vor}    
\end{figure}

Figure \ref{fig:vor} shows the perturbed gas vortensity $\delta q$, advection term $s'_\mathrm{adv}=-\delta\vgx (\partial q_0/\partial x)/(-i\omega + i\ky v_\mathrm{gy0})$, and drag force term $s'_\mathrm{drag}=S'_\mathrm{drag}/(-i\omega + i\ky v_\mathrm{gy0})$ in $(x-y)$ plane. The top panels of Figures \ref{fig:vorI} and \ref{fig:vorII} show the same quantities at $y=0$.
For the Type I mode, a pair of classic Rossby waves dominates the perturbed vortensity. The advection term is the major contribution to the perturbed vortensity, consistent with the instability mechanism of the pure gas RWI. In Figure \ref{fig:vorI}, the drag force term exhibits an opposite sign relative to $\delta q$ for one of the Rossby waves, indicating that dust acts to damp the Type I DRWI modes.

The right column of Figure \ref{fig:vor} shows the left mode of a pair of Type II modes, presented in Figure \ref{fig:efII}. In addition to the classic Rossby waves, a pair of thin waves is present between the classic waves, which is the key distinction from Type I modes. The advection term drives the classic Rossby waves, while the drag force term serves as the underlying source of the thin waves. The amplitude of the perturbed vortensity $\delta q$ of the thin waves is comparable to that of the classic Rossby waves. In Figure \ref{fig:vorII}, the drag force term has the same sign as $\delta q$, which acts to promote the Type II modes. 

\begin{figure}
    \centering
    \includegraphics[width=0.5\textwidth]{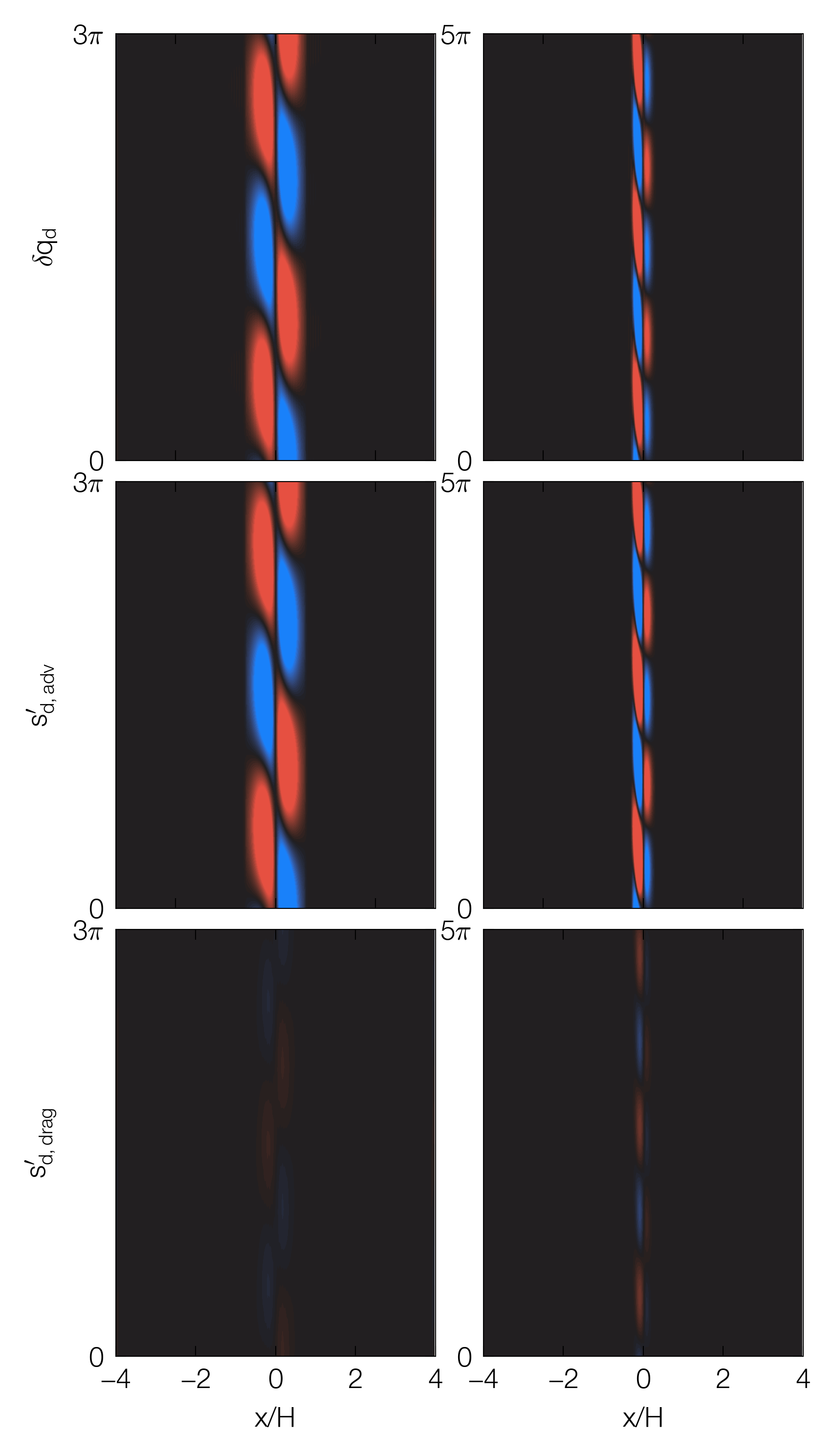}
    \caption{Same to Figure \ref{fig:vor} but with terms in the perturbed dust vortensity equation (\ref{eq:pvd}).
    }
    \label{fig:vord}    
\end{figure}

We also analyze the dust vortensity budget. The perturbed dust vortensity equation is (Appendix \ref{app:vor})
\begin{equation}
(-i\omega + i\ky v_\mathrm{dy0})\delta q_\mathrm{d} + \delta\vdx \frac{\p q_\mathrm{d0}}{\p x} \approx S'_\mathrm{d,drag},
\label{eq:pvd} 
\end{equation}
where the drag force term dominates over the dust diffusion term for $S'_\mathrm{d}$. The perturbed dust vortensity driven by the drag force is expressed by 
\begin{equation}
S'_\mathrm{d,drag} = -\frac{1}{\rd\ts}\nabla\times(\delta\vd-\delta\vg). 
\end{equation}
Figure \ref{fig:vord} shows the dust perturbed vortensity $\delta q_\mathrm{d}$, advection term $s'_\mathrm{d,adv}=-\delta\vdx (\partial q_\mathrm{d0}/\partial x)/(-i\omega + i\ky v_\mathrm{dy0})$, and drag force term $s'_\mathrm{d,drag}=S'_\mathrm{d,drag}/(-i\omega + i\ky v_\mathrm{dy0})$ in $(x-y)$ plane. 
For both Type I and Type II modes, a pair of thin waves, corresponding to the background dust width, emerge on both sides of the corotation radius. The advection term is clearly the dominant contributor to the perturbed dust vortensity in both cases. This contrasts with the perturbed gas vortensity shown in Figure \ref{fig:vor}, where the drag force serves as the source of thin waves.

\subsection{Parameter Study}\label{sec:ps}

We perform parameter study on the growth rates of Type I and Type II DRWI. The free parameters in the linear analysis include the gas bump amplitude $A$, gas bump width $w$, dust bump amplitude $A_d$, dust bump width $w_d$, Stokes number $\mathrm{St}$, and viscosity $\alpha$.
The parameter ranges are chosen to be gas bump amplitude $A \in [1.2, 3.8]$, dust bump amplitude $A_d/A = \{0.1, 0.3\}$, gas bump width $w/H = \{1, 1.2, 1.5\}$, and dust bump width $w_d/H = \{0.1, 0.3\}$. The Stokes numbers are taken to be $\mathrm{St} = \{ 0.01, 0.1\}$. The viscous parameter is set to $\alpha = \{10^{-4}, 5 \times 10^{-4}\}$, consistent with both theoretical predictions and observational data \citep{flaherty_etal17, flaherty_etal18, flaherty_etal20, lesur+23}. Figures \ref{fig:para1}-\ref{fig:para4} show the growth rates across our parameter space. 

\begin{figure}
    \centering
    \includegraphics[width=0.5\textwidth]{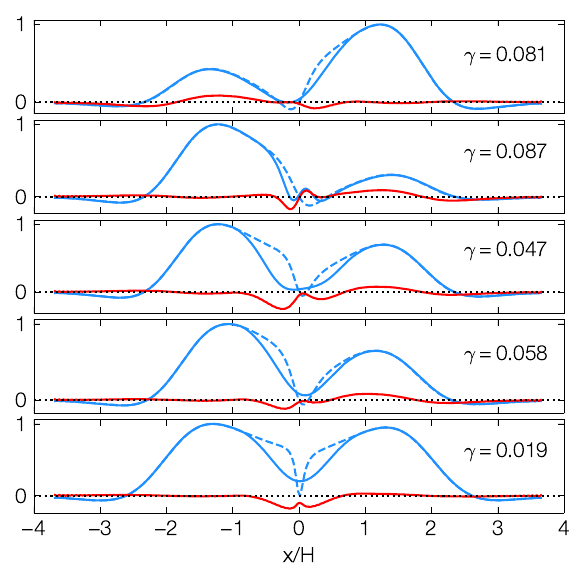}
    \caption{
    Real part of $\delta q$ (solid blue), advection term  $s'_\mathrm{adv}$ (dashed blue), and drag force term  $s'_\mathrm{drag}$ (red) at $y=0$, normalized by the maximum of real $\delta q$.
    Top panel: Type I mode in Figure \ref{fig:vor}, with $A=1.5$, $w=H$,  $\epsilon_\mathrm{max}=0.1$, $w_d=0.3$.
    Lower panels have almost identical parameters, but with one parameter changed in each. Specifically, the changes are $w_d = 0.1$, $\epsilon_\mathrm{max} = 0.3$, $A = 1.2$, and $w/H = 1.1$, from top to bottom, respectively.
    }
    \label{fig:vorI}    
\end{figure}

For Type I modes, the growth rates decrease monotonically with $A_d$ and $w_d$ (Figure \ref{fig:alp1e-3}, Figures \ref{fig:para1}-\ref{fig:para4}). When varying the gas bump parameters, the growth rates increase with $A$ but drop significantly with $w$. Notably, when $A_d/A = 0.3$, a slight increase in $w/H$ from 1 to 1.2 can completely suppress the Type I modes. 
These trends can be explained by the impact of drag force on the perturbed vortensity.
In Figure \ref{fig:vorI}, we see that larger values of $A_d$, $w_d$, and $w$, or smaller values of $A$, result in a greater suppression of $\delta q$ in the inner part of the classic Rossby waves, due to the drag force term (solid red).

\begin{figure}
    \centering
    \includegraphics[width=0.5\textwidth]{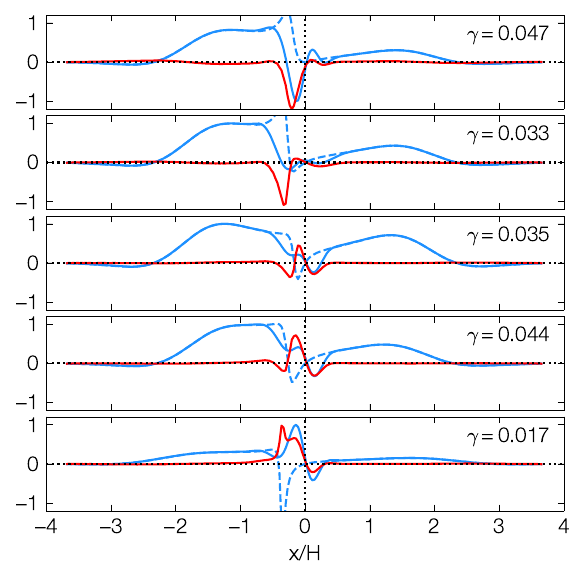}
    \caption{
    Same as Figure \ref{fig:vorI} but for Type II modes.    
    Top panel: Type II mode in Figure \ref{fig:vor}, with $A=1.5$, $w=H$,  $\epsilon_\mathrm{max}=0.5$, $w_d=0.1$.
    Lower panels have almost identical parameters, but with one parameter changed in each. Specifically, the changes are $w_d = 0.2$, $\epsilon_\mathrm{max} = 0.3$, $A = 1.2$, and $w/H = 1.2$, from top to bottom, respectively.
    }
    \label{fig:vorII}    
\end{figure}

\begin{figure*}
    \centering
    \includegraphics[width=1\textwidth]{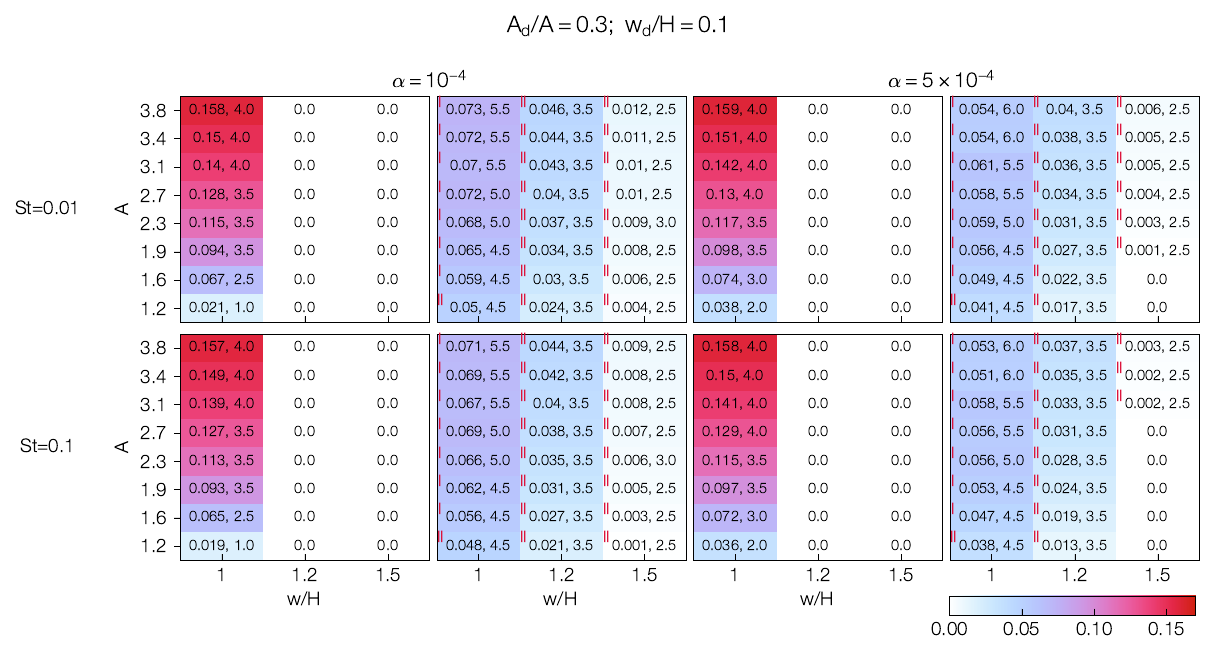}
    \caption{
    Maximum growth rates (first numbers in the cell) and the corresponding azimuthal wave numbers $\ky$ (second numbers in the cell) as a function of $A$ and $w$ at $A_d/A=0.3$ and $w_d/H=0.1$.
    Color in each cell denotes growth rates $\gamma/\Omega $. Zero denotes no modes are found for $\ky\in[1,7]$. 
    First row: $\mathrm{St=0.01}$, second row: $\mathrm{St=0.1}$. 
    First and third columns: growth rates of Type I modes. Second and forth columns: growth rates of Type II modes. 
    First and second columns: $\alpha=10^{-4}$. 
    Third and forth columns: $\alpha=5\times 10^{-4}$. 
    Red text `I' or `II' in the upper left corner of each cell denotes which Type of DRWI has higher growth rate (Type I or Type II). 
    }
    \label{fig:para1}    
\end{figure*}

\begin{figure*}
    \centering
    \includegraphics[width=1\textwidth]{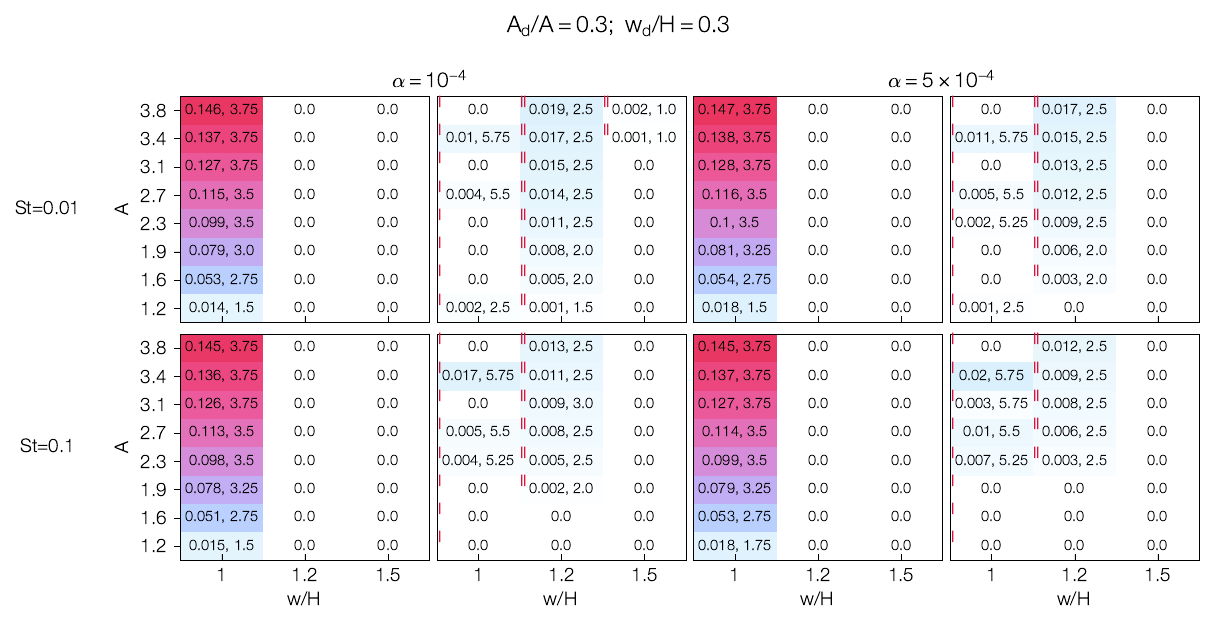}
    \caption{Same as Figure \ref{fig:para1} with $A_d/A=0.3$ and $w_d/H=0.3$.    
    }
    \label{fig:para2}    
\end{figure*}

\begin{figure*}
    \centering
    \includegraphics[width=1\textwidth]{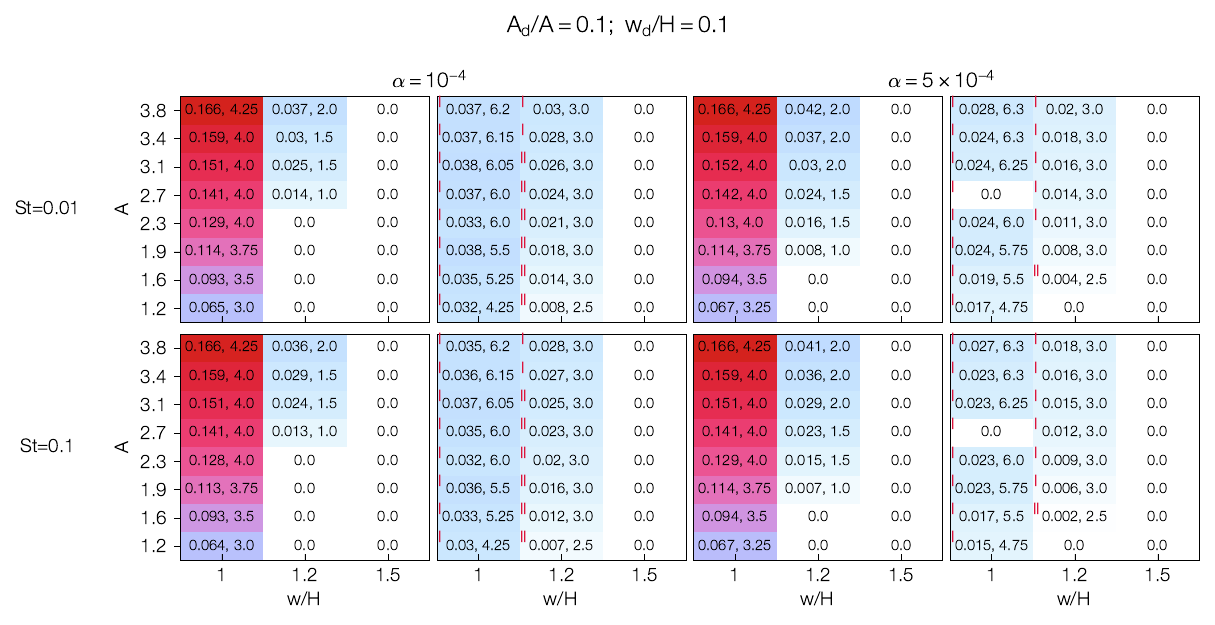}
    \caption{Same as Figure \ref{fig:para1} with $A_d/A=0.1$ and $w_d/H=0.1$.  
    }
    \label{fig:para3}    
\end{figure*}

\begin{figure*}
    \centering
    \includegraphics[width=1\textwidth]{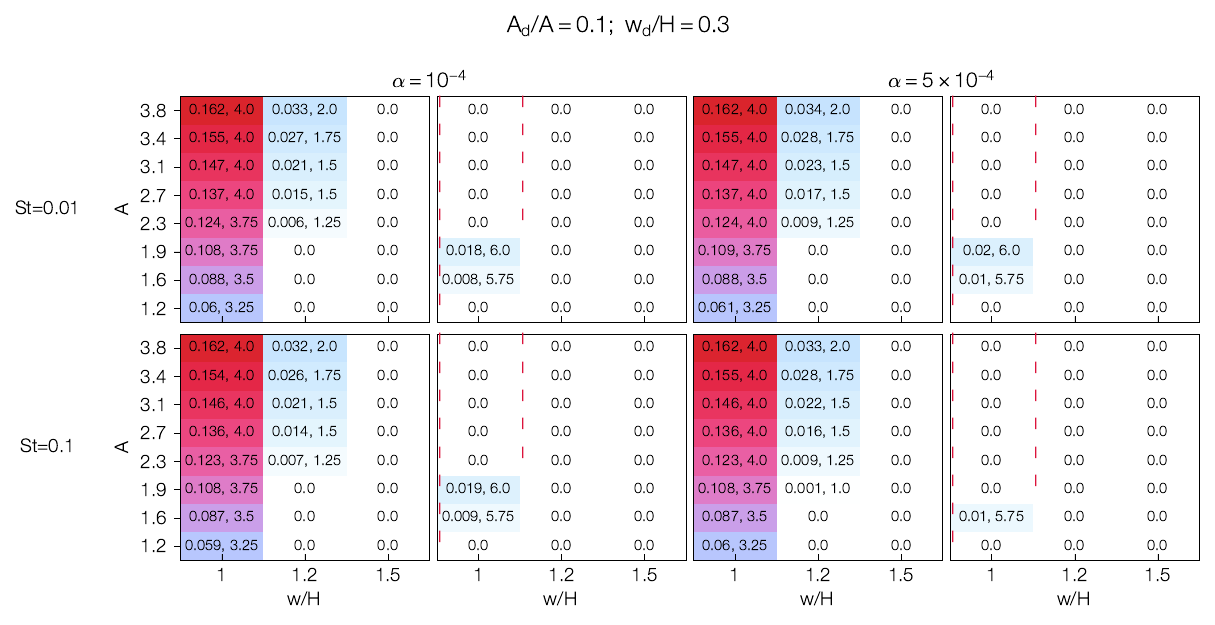}
    \caption{Same as Figure \ref{fig:para1} with $A_d/A=0.1$ and $w_d/H=0.3$.  
    }
    \label{fig:para4}    
\end{figure*}

For Type II modes, the growth rates increase monotonically with $A_d$, in contrast to Type I modes (Figure \ref{fig:alp1e-3}, Figures \ref{fig:para1}-\ref{fig:para4}). The growth rates do not vary monotonically with $w_d$, as shown in Figure \ref{fig:alp1e-3}, where maxima occur at $w_d = 0.1$ but vanish at $w_d = 0.3$. When varying the gas bump parameters, the growth rates increase with $A$, but decrease with $w$, similar to Type I modes.
These trends can also be attributed to the effect of drag force term on the perturbed vortensity. Type II modes rely on the presence of thin waves in the perturbed vortensity. In Figure \ref{fig:vorII}, smaller values of $A_d$ and $A$, or larger values of $w_d$ and $w$, result in weaker thin Rossby waves due to the drag force term (solid red).

Moreover, the growth rates show a slight increase with larger $\alpha$ for Type I modes. The underlying reason for this trend is not yet fully understood and warrants further investigation in future work. The Stokes numbers of $\mathrm{St} = 0.01$ and $\mathrm{St} = 0.1$ do not produce significant differences in growth rates of Type I modes, consistent with the results shown in Figure \ref{fig:alp1e-3}.
The growth rates decrease with $\alpha$ for Type II modes. The Stokes number of $\mathrm{St=0.01}$ gives rise to slightly stronger growth rates than $\mathrm{St=0.1}$, as Type II modes prefer stronger coupling between gas and dust. 

Local numerical simulations revealed that the non-linear evolution of Type II modes can yield ring-like substructures, whereas Type I modes gave rise to vortices \citepalias{lb23}. Thus, we search the parameter space where Type II growth rates are greater than Type I.
When $A_d/A=0.3$ (Figures \ref{fig:para1} and \ref{fig:para2}), Type II modes tend to dominate generically for $w/H$ greater than unity, because Type I modes are suppressed.
When $A_d/A=0.1$ (Figures \ref{fig:para3} and \ref{fig:para4}), Type I modes dominate in most cases because the dust content is low. Type II modes only take over for $w/H$ greater than unity and small $A$ when $w_d/H=0.1$. When $w_d/H=0.3$, most of the Type II modes vanish.

\subsection{Comparison to \citetalias{lb23}}\label{sec:comp}

\citetalias{lb23} were the first to identify DRWI via linear theory and shearing box numerical simulations. Although all terms in the conservation laws of equations \eqref{eq:2}-\eqref{eq:4} are identical between their work and ours, they used single-fluid formulation to conduct linear analysis, combining dust and gas equations \citep{ly17}. In contrast, we adopt the two-fluid equations, as applied in the linear analysis of streaming instability \citep{yg05}, where dust and gas are treated as separate fluids. As a result, there are a number of similarities and differences between our work and theirs. For clarity, we highlight the key points below.

First, we discuss the similarities. Both works identify two distinct DRWI modes (Type I and II). This is achieved by setting the equilibrium radial velocities to zero, either for the single-fluid model in \citetalias{lb23} or for the dust and gas components in this work, which effectively filters out the streaming instability. Type I modes are closely related to the pure gas RWI, whereas Type II modes arise from dust-gas coupling. Additionally, the oscillation frequencies $w_r$ of Type I modes approach zero (but not equal to), while those of Type II modes are generally greater. For Type I modes, the perturbed vortensity is primarily driven by the advection of background vortensity, whereas for Type II modes (thin waves), the perturbation is driven by dust-gas drag (baroclinic term in \citetalias{lb23} as they combined the two-fluid equations).

The major distinction between this work and \citetalias{lb23} lies in the behavior of the two types of DRWI modes. \citetalias{lb23} found that both modes can coexist for a given 
$\ky$ (see their Appendix D), whereas our results indicate that for a given $\ky$, only one mode grows at a time. However, as $\ky$ varies, our results suggest that a transition between the two types of modes can occur. The underlying cause of this discrepancy remains unclear. Future studies could further investigate this issue, possibly by exploring alternative dust diffusion models and comparing the results.

\section{DRWI in ALMA Rings}\label{sec:ALMA}

We investigate the onset of DRWI within the parameter space spanned by ALMA rings of DSHARP sources. In the following subsections, we discuss the bump parameters constrained by observations (\S\ref{sec:para}), the presence of DRWI and its implication to dust growth in rings (\S\ref{sec:ex}).

This paper focuses on the linear behavior of the DRWI. However, the linear phase is usually transient, and to facilitate direct comparison with ALMA rings, it is essential to examine the nonlinear evolution of the instability. Local shearing-box simulations on (radial) $H$-scales have shown that Type II modes can yield ring-like substructures, while Type I modes tend to produce vortices \citepalias{lb23}.
Given the tendency of Type I modes to generate non-axisymmetric features, we expect most ALMA rings to be stable or only marginally unstable to Type I DRWI. This expectation is consistent with our findings for the DSHARP sources. On the other hand, ring-like structures persist under Type II DRWI in local simulations. Whether these structures can survive in global simulations remains an open question, and future numerical studies are needed to explore this further. Detailed discussions follow below.

\subsection{Gaussian Bump Parameters}\label{sec:para}

\begin{table*}
\caption{Parameters for Rings in DSHARP Sources.} 
\begin{tabular*}{1\textwidth}{l@{\hspace{1.1cm}}l@{\hspace{1.1cm}}l@{\hspace{1.1cm}}l@{\hspace{1.1cm}}l@{\hspace{1.1cm}}l@{\hspace{1.1cm}}l@{\hspace{1.1cm}}l@{\hspace{1.1cm}}
}
\hline
\hline
Source & Ring & Name & $A_d$ & $w_d/H$ & $w_\mathrm{min}/H$ & $w_\mathrm{max}/H$ & $w_\mathrm{rot}/H$   \\ 
(1) & (2) & (3) & (4) & (5) & (6) & (7) & (8)  \\ 
\hline 
AS 209 & 1 & B74 & 0.19 & 0.6 & 1 & 3.5& 1.4  \\
AS 209 & 2 & B120 & 0.2 & 0.4 & 1 & 2& 1   \\
Elias 24  & 1  & B77 & 0.17 & 0.6 & 1 & 2& /  \\
HD 163296 & 1 & B67  & 0.19 & 1.6 & 1.6 & 3& 3.3  \\
HD 163296 & 2  & B100 & 0.14 & 0.7 & 1& 2& 3.5  \\
GW Lup & 1 & B85 & 0.13 & 0.6 & 1 & 1.25& /  \\ 
HD 143006 & 1 & B41 & 0.1 & 1.9 & 1.9& 5& / \\
HD 143006 & 2 & B65 & 0.09 & 2 & 2& 2.7& / \\  
\hline 
\hline 
Fiducial &  & & $A$ & & $w/H$ &  $\alpha$   \\
 &  & & (9) & & (10) &  (11)  \\ 
\hline 
 &  &  & 1.5 &  & 1 & $10^{-3}$ \\ 
\hline \\
\end{tabular*}
\label{table:t1} 
{\raggedright 
Note. (1) Subsample of DSHARP sources. (2) Internal numbering of rings in \citetalias{dullemond+18}. (3) Ring name from \citet{huang_etal18}. (4) Dust bump amplitude. (5) Dust bump width. (6) Minimum gas bump width. (7) Maximum gas bump width. (8) Gas bump width by rotation curve. (9) Fiducial gas bump amplitude (\S\ref{sec:re}). (10) Fiducial gas bump width. (11) Fiducial viscous $\alpha$ value. See \S\ref{sec:para} for more detailed explanation. 
\par}
\end{table*}

\begin{table*}
\caption{Excitation of DRWI at $\mathrm{St=10^{-3}}$, $\alpha=10^{-4}$ and $\ky\in[1,7]$.} 
\begin{tabular*}{1\textwidth}{l@{\hspace{1.4cm}}l@{\hspace{1.4cm}}l@{\hspace{1.4cm}}l@{\hspace{1.4cm}}l@{\hspace{0.8cm}}l@{\hspace{0.8cm}}l@{\hspace{0.8cm}}
}
\hline
\hline
Source & Ring & Name & $A_d/A$ & Type I; Type II  & Type I; Type II  &  Type I; Type II  \\ 
(1) & (2) & (3) & (4) & (5) & (6) & (7)  \\ 
 &  &  &  & $w_\mathrm{min}$ & $w_\mathrm{max}$ & $w_\mathrm{rot}$ \\ 
\hline 
AS 209 & 1 & B74  & 0.1 & Y; N & N; N & N; N  \\
AS 209 & 2 & B120 & 0.1 & Y; N & N; N & Y; N  \\
Elias 24  & 1  & B77 & 0.1 & Y; N & N; N & /  \\
HD 163296 & 1 & B67  & 0.1 &  N; N &  N; N &  N; N \\
HD 163296 & 2  & B100  & 0.1 & Y; N & N; N & N; N\\
GW Lup & 1 & B85 & 0.1 & Y; N &  N; N & /\\ 
HD 143006 & 1 & B41 & 0.1 &  N; N &  N; N & / \\
HD 143006 & 2 & B65 & 0.1 & N; N &  N; N & / \\
\hline \\
\end{tabular*}
\label{table:t2} 
{\raggedright 
Note. (1), (2) and (3) are same as Table \ref{table:t1}. (4) dust-to-gas ratio at bump center $\epsilon_\mathrm{max}=A_d/A$. (5) onset of DRWI modes (Type I and II) for $w_\mathrm{min}$ in Table \ref{table:t1}. (6) onset of DRWI modes for $w_\mathrm{max}$ in Table \ref{table:t1}. (7) onset of DRWI modes for $w_\mathrm{rot}$ in Table \ref{table:t1}. 
\par}
\end{table*}

The dust and gas bump parameters, $w_d,w,A_d,A$, can be either observationally constrained, thanks to the DSHARP Programme (\citealp{andrews+18}, \citetalias{dullemond+18}), or approximated by numerical simulations of planet-disk interactions. We detail the values chosen below.

\citetalias{dullemond+18} selected a sample of sufficiently face-on DSHARP sources, which possess radially well separated, high-contrast dust rings. The dust bump width $w_d$ is directly obtained by fitting a Gaussian profile to the observed ring emission. \citetalias{dullemond+18} employed a Gaussian function similar to our Eq. \eqref{eq:rd0}, but without the density floor that applied far from the bump center. Their fittings gave $w_d/H$ for eight rings, listed in Table \ref{table:t1}. 
The gas bump width $w$ is constrained by \citetalias{dullemond+18} as well. A lower limit is set by $w_\mathrm{min}=w_d$, and an upper limit $w_\mathrm{max}$ is set by the separation of the ring and the nearest intensity minimum. 
\citet{rosotti+20} further estimated gas bump width $w_\mathrm{rot}$ by extracting the gas rotation curve from the emission lines, for four rings in HD 163296 and AS 209, listed in Table \ref{table:t1}.

Moreover, the dust bump amplitude $A_d$ can also be directly obtained from the Gaussian fitting of intensity. The dust density amplitude derived from deconvolved dust intensity is listed in Table \ref{table:t1}, using eq. 7 in \citetalias{dullemond+18} for which thermal emission is assumed optically thin.

To estimate the gas bump amplitude $A$, constraints shall be placed on the dust-to-gas density ratio $\Sigma_d/\Sigma_g$ at the ring center. \citetalias{dullemond+18} estimated the dust surface density $\Sigma_d$ by assuming optically thin thermal emission. They further place an upper limit on the gas surface density $\Sigma_{g, \mathrm{max}}$ by demanding the disk to be gravitationally-stable (Toomre $Q>2$; \citet{toomre64}). Unfortunately, this upper limit can only give a lower limit on dust-to-gas mass ratio $\sim0.01$. 
We then opt for determine $\Sigma_d/\Sigma_g$ from numerical simulations, using those with planets carving gaps and rings. These simulations found $\Sigma_d/\Sigma_g$ between a few times of $0.01$ and $0.1$ \citep[e.g.][]{Paar04,pinilla+12,Dra+19}. Thus, we take $\Sigma_d/\Sigma_g=A_d/A=0.1$ at the bump center.

Note that the unstratified shearing box employed throughout this work is applicable to the midplane of the disk, where vertical gravity is nearly negligible. Above, we use surface densities to determine the dust-to-gas ratio, which is justified since both gas and dust in protoplanetary disks typically follow Gaussian distributions vertically, with mass densities peaking at the midplane.

The Stokes number is chosen to be $\mathrm{St = 10^{-3}}$, as observational constraints suggest it typically ranges from $10^{-3}$ to $10^{-2}$ (\citetalias{dullemond+18}, \citealp{rosotti+20}). In Figure \ref{fig:alp1e-3}, we find that $\mathrm{St = 10^{-3}}$ and $\mathrm{St = 10^{-2}}$ produce similar growth rates. The viscosity is selected to be relatively low, with $\alpha = 10^{-4}$ \citep{flaherty_etal17, flaherty_etal18, flaherty_etal20, rosotti+20}.

\subsection{Excitation of DRWI in Rings}\label{sec:ex}

It is found that while Type I modes can be triggered for $w_\mathrm{min}$ in some rings, they do not emerge for $w_\mathrm{max}$. When the rotation curve provides a measurement of $w_\mathrm{rot}$, only one out of four rings meets the conditions for the onset of Type I mode. Local numerical simulations by \citetalias{lb23} showed that the non-linear stage of Type I modes can lead to azimuthal clumps, similar to the pure gas RWI. The absence of Type I modes in many ALMA rings listed in Table \ref{table:t1} suggests that only a limited number of pressure bumps shaped by annular substructures are capable of inducing azimuthal asymmetries. This is in alignment with the ALMA observations, which have showed that only around 10\% of the ringed disks exhibit crescent-like azimuthal asymmetries \citep{huang_etal18}.

%Moreover, the optical depth can make an intrinsic narrower dust ring with the coagulation-fragmentation equilibrium to be responsible for a wider ring observed at mm band. The reason is as follows. After considering the equilibrium of dust coagulation and fragmentation, it has been suggested that the intrinsic dust ring for different species follows the same spatial distribution with a width of the order of $0.1-0.2\ w_{\rm g}$ \citep{Yang+25}. As the small dust species can contribute a large fraction of opacity at mm band, it is very likely that the emission at the mm band is optically thick. Such a large opacity from all dust species thus results in a dust ring looking wider in mm observations. Applying this to the multi-wavelength observations of HD 163296, it is found that the intrinsic dust ring width at 67 au and 100 au is indeed a factor of two narrower than what we observed.

Type II modes are completely absent in all eight rings, as the dust bump width $w_d$ is too large to excite these modes. 
Taking Ring 1 (B74) of AS 209 as an example, $w_d/H$ needs to be as small as $0.1$ or lower to excite the Type II DRWI using $w_\mathrm{rot}$.
However, it could be that narrower rings are too thin to be observed by ALMA. For example, \citet{jennings+22} applied the super-resolution code FRANK to 20 DSHARP sources \citep{jennings+20} and found that high-contrast rings are brighter and on average 26\% narrower. Hence, the measurements of ring widths of DSHARP sources should be treated as upper limits due to the limited spatial resolution. 
Moreover, the intrinsic dust ring widths can be different from the observed values, due to the optically thin assumption made in \citetalias{dullemond+18}. The intrinsic optical depth might be greater as small dust species can contribute a large fraction of opacity at mm band, and the emission becomes optically thick. This can lead to the observed rings looking wider at ALMA band.  
The coagulation-fragmentation processes can modify the dust size distribution within the ring, and may further enhance the optical depth \citep{Yang+25}. This enhancement could make the observed rings even wider. 
If many rings are intrinsically considerably narrower compared to DSHARP observations, Type II DRWI may still be plausible to operate. 

Shearing box simulations by \citetalias{lb23} demonstrated that Type II modes can lead to the persistence of rings and the formation of small gravitationally bound clumps in the nonlinear stage. However, these simulations were performed on extremely narrow radial domains (on $H-$scales), limiting their applicability to global disk structures.
Larger-scale two-fluid simulations have been carried out in the context of a Neptune-mass planet carving rings \citep{cp24}, but this study did not directly compare their results with the linear theory of DRWI. Future global simulations with high spatial resolution are needed to validate or revise the conclusions drawn from shearing box simulations.

Another possibility is that the observed rings are already the outcome of the nonlinear evolution of Type II modes. If so, this would suggest that while Type II modes can support the persistence of rings, they are not efficient at triggering Type I modes and producing azimuthal asymmetries.

Finally, it is also possible that Type II DRWI is absent in many observed rings, implying that additional physical processes may be required to promote dust growth and planetesimal formation in annular substructures.
A companion paper in this series investigates the linear behavior of the global SI in the presence of a pressure bump. The key question is whether SI can operate within a pressure bump and how its growth depends on bump parameters in both the linear and nonlinear regimes.

%If the eight rings in \citetalias{dullemond+18} are representative, this suggests that Type II DRWI is unlikely to be the primary mechanism for forming annular substructures.

\section{Conclusions}\label{sec:c}

In this paper, we investigate the linear properties of the DRWI in the presence of a radial pressure bump, representing annular substructures in protoplanetary disks. To this end, we perform a linear analysis of the two-fluid dust–gas equations, setting the background radial velocities to zero to suppress the SI. The spectral code \textsc{Dedalus} is employed to solve the eigenvalue problem. Our main findings are summarized below:

\begin{enumerate}
    \item Two types of DRWI modes are identified: Type I modes, which are dust-modified versions of the classic gas RWI, and Type II modes, which arise from dust-gas coupling. This confirms the findings of \citetalias{lb23}, who used a single-fluid formulation for dust and gas.
    \item Type I and Type II modes never coexist for a given $\ky$. Typically, Type I modes dominate at lower $\ky$, while Type II modes emerge at higher $\ky$. A transition between the two occurs as $\ky$ varies.
    \item The instability mechanism of Type I modes is driven by the advection of background vorticity into perturbed vorticity. Type II modes consist of two main waves: the classic Rossby waves, which are advection-driven, and the thin waves, which are driven by dust-gas drag.
    \item The growth rates of Type I (Type II) modes decrease (increase) with dust-to-gas ratio $\epsilon_\mathrm{max}$ (or equivalently, $A_d$). For both types, the growth rates are enhanced by increasing $A$ or decreasing $w$.
    \item Applying bump parameters constrained from ALMA rings (DSHARP sources), we find that Type I modes can emerge in one out of four rings, potentially explaining the general absence of azimuthal asymmetries in many ALMA disks.
    \item 
    Type II modes are absent in all eight ALMA rings. However, it remains an open question whether narrower rings, capable of triggering Type II DRWI, are more common in protoplanetary disks but remain unresolved by ALMA, or ALMA rings appear artificially widened due to optical depth effects. Alternatively, the absence of Type II DRWI in these systems may be genuine, suggesting that additional mechanisms are needed to promote dust growth and planetesimal formation within annular substructures.
    
\end{enumerate}

%%%%%%%%%%%%%%%%% Acknowledgements %%%%%%%%%%%%%%%%%%%%%

\section*{Acknowledgements}

We thank our referee, Henrik Latter, for the detailed comments that improved the clarity of this manuscript.
We thank Hanpu Liu and Lile Wang for the fruitful discussions. 
CC acknowledges funding from NSERC Canada and Nanjing University.
YPL is supported in part by the Natural Science Foundation of China (grants 12373070 and 12192223), the Natural Science Foundation of Shanghai (grant NO. 23ZR1473700). 
ZX acknowledges funding from the Carlsberg Foundation (Semper Ardens: Advance grant FIRSTATMO).
RL acknowledges support from the Heising-Simons Foundation 51 Pegasi b Fellowship.
CY is supported by the National SKA Program of China (grant 2022SKA0120101) and the National Natural Science Foundation of China (grants 11873103 and 12373071). 
MKL is supported by the National Science and Technology Council (grants 112-2112-M-001-064-, 113-2124-M-002-003-) and an Academia Sinica Career Development Award (AS-CDA-110-M06). 

\section*{Data Availability}

The data underlying this article will be shared on reasonable request to the corresponding author.

\bibliography{disk}{}
\bibliographystyle{aasjournal}

%%%%%%%%%%%%%%%%% APPENDICES %%%%%%%%%%%%%%%%%%%%%
\appendix
% \onecolumn

\section{steady state dust density}\label{app:rhod}

In \S\ref{sec:nds}, we derive the steady state solutions in the non-drift regime. Given $v_\mathrm{gy0}=v_\mathrm{dy0}$, it requests that the dust diffusion term in equation \eqref{eq:A5} and the gas pressure gradient in equation \eqref{eq:A3} to be equal. This yields the following second order ordinary differential equation,
\begin{equation}
\frac{1}{\rho_\mathrm{d0}}\frac{\p}{\p x}[\rho_\mathrm{d0} D_\mathrm{d}^2 (\p\ln \epsilon_0/\p x)^2] +\frac{1}{\rho_\mathrm{g0}}\frac{\p P_0}{\p x}=0.
\label{eq:a1}
\end{equation}
We solve \ref{eq:a1} numerically as an initial value problem. We first convert it to two first order ODEs, by defining 
\begin{equation}
y_1 =\epsilon_0,
\end{equation}
\begin{equation}
y_2 = \frac{\p\ln\epsilon_0}{\p x}.
\end{equation}
Then, \ref{eq:a1} is cast into
\begin{equation}
dy_1 = y_1y_2,
\label{eq:a4}
\end{equation}
\begin{equation}
dy_2 = -\frac{1}{2}\bigg[y_2^2+y_2\frac{\p\ln\rho_\mathrm{g0}}{\p x} +\frac{1}{y_2}\frac{1}{D_\mathrm{d}^2\rho_\mathrm{g0}}\frac{\p P_0}{\p x} \bigg].
\label{eq:a5}
\end{equation}
We expect a Gaussian bump for dust density. This agrees with the continuum observations of rings and gaps in ALMA \citep{huang_etal18,andrews20}. Hence, the initial conditions are given by
\begin{equation}
y_1(x=0) = \epsilon_\mathrm{max},
\end{equation}
\begin{equation}
y_2(x=0) = 0.
\end{equation}
In practive, we note that $y_2$ appears in the denominator of \ref{eq:a5}. To avoid the singularity at $x=0$, we opt for integrating the ODEs in the domain of $x\in[x_\mathrm{min},4H]$, where $x_\mathrm{min}=10^{-10}H$. To find $y_2$ at $x=x_\mathrm{min}$, we take the limit of $x\rightarrow 0$ of \ref{eq:a5}. With L'Hospital's rule, we arrive at  
\begin{equation}
\frac{dy_2}{dx}\bigg|_{x=0} = -\frac{1}{2D_\mathrm{d}(1+A)}\frac{\frac{d^2P_0}{dx^2}\big|_{x=0}}{\frac{dy_2}{dx}\big|_{x=0}}.
\end{equation}
This is a quadratic equation in $\frac{dy_2}{dx}\big|_{x=0}$, and we take the negative root as at $x=0$, $\epsilon=\epsilon_\mathrm{max}$.
Thus, the boundary conditions at $x=x_\mathrm{min}$ are 
\begin{equation}
y_1(x=x_\mathrm{min})=\epsilon_\mathrm{max},
\end{equation}
\begin{equation}
y_2(x=x_\mathrm{min})=x_\mathrm{min}\frac{dy_2}{dx}\bigg|_{x=0}.
\end{equation}
\ref{eq:a4} and \ref{eq:a5} are solved by \texttt{scipy.integrate.odeint}, which uses LSODA from the FORTRAN library odepack.

%\section{Linearized dust diffusion terms}\label{app:le}

\section{Boundary conditions and Fourier basis}\label{app:bc}

Density waves can appear in the eigenfunctions of (D)RWI. They are launched at the Lindblad resonances and propagate away from the pressure bump \citep{li_etal00}. Previous work on the linear analysis of RWI typically employed boundary conditions to accommodate density waves (\citealp{ono_etal16}, \citetalias{lb23}). This is achieved by applying WKB analysis in $x$ at the boundaries, where the steady-state quantities vary slowly, and the eigenfunctions are expressed as $\propto\exp(ik_\mathrm{x}x+i\ky y-i\omega t)$. However, this approach involves the computation of $k_\mathrm{x}$ at the boundaries, adding complexity to solving the EVPs (see Appendix A2 of \citetalias{lb23}). 
In this work, we adopt the Fourier basis, which naturally satisfies periodic boundary conditions. This choice is motivated by its simplicity, as there is no need to compute $k_\mathrm{x}$ at the boundaries. Below, using the numerical method described in \S\ref{sec:me}, we demonstrate that 1) the Fourier basis effectively captures density waves, and 2) whether or not density waves are included in the domain has little impact on the growth rates of (D)RWI.

\begin{figure}
    \centering
    \includegraphics[width=0.5\textwidth]{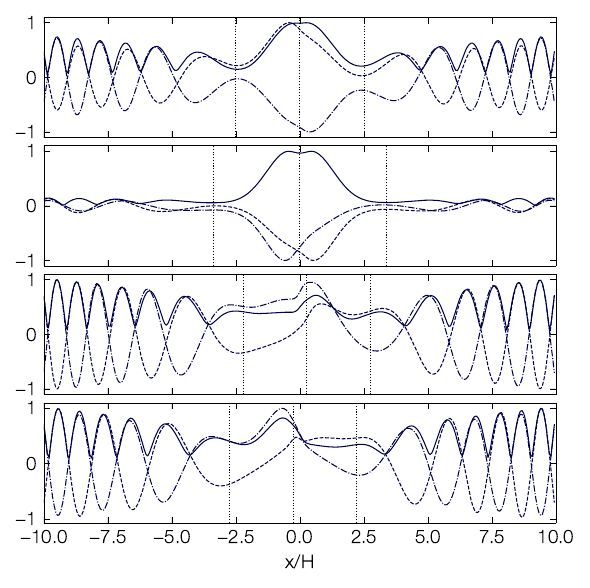}
    \caption{Normalized real (dashed), imaginary (dash-dotted), and absolute (solid) part of the perturbed gas density in the domain of $x/H\in[-10,10]$. 
    The vertical blue lines show the corotation, inner and outer Lindblad radii.
    First row: gas RWI with parameters $H/r_0=0.1$, $A=1.5$, $w=H$, $\alpha=10^{-3}$, $k_yr_0=3$, $x_\mathrm{c}/H=-0.022$, $x_\mathrm{ILR}/H=-2.505$, $x_\mathrm{OLR}/H=2.505$. 
    Second row: Type I RWI mode with parameters $k_yr_0=2$, $\epsilon=0.5$, $w_d=0.05$, $\mathrm{St=0.01}$, $x_\mathrm{c}/H=-0.0285$, $x_\mathrm{ILR}/H=-3.39$, $x_\mathrm{OLR}/H=3.39$.
    Third and fourth rows: a pair of Type II RWI modes with parameters $k_yr_0=3$, $\epsilon=0.5$, $w_d=0.3$, $\mathrm{St=0.01}$, $x_\mathrm{c}/H=0.256, -0.256$, $x_\mathrm{ILR}/H=-2.193, 2.193$, $x_\mathrm{OLR}/H=2.764, -2.764$.
    }
    \label{fig:dw}    
\end{figure}

Figure \ref{fig:dw} shows the normalized perturbed gas density $\delta\rg$, for the pure gas RWI, Type I, and a pair of Type II modes, respectively (from top to bottom). 
The corotation radius $x_\mathrm{c}$, where $Re[\Delta\omega]=0$ and the inner and outer Lindblad radii $x_\mathrm{ILR}$, $x_\mathrm{OLR}$, where $\Delta\omega^2=\kappa^2$ and $\kappa^2=2\Omega (2\Omega +\p v_\mathrm{gy0}/\p x)$, are shown as vertical lines. 
The radial domain is extended to $x/H\in[-10,10]$ in Figure \ref{fig:dw} in order to accommodate the density waves. It is clear that in all the panels density waves are captured. The growth rates from top to bottom panel are $\gamma/\Omega =0.039, 0.084, 0.037, 0.037$, repsectively.
Now, we test with the domain of $x/H\in[-4,4]$ employed in the main text. Density waves do not propagate far, yet the growth rates remain largely unaffected, with $\gamma/\Omega =0.041, 0.083, 0.039, 0.039$, respectively. The reason behind is (D)RWI structure is concentrated around $x=0$, whereas density waves emerge farther from the domain center, and hence have little impact on the (D)RWI.

\section{Vortensity equation}\label{app:vor}

To understand the DRWI mechanism, we shall investigate how the background vortensity is transferred to Rossby waves. This is done by deriving the vortensity equation. 
Taking the curl of eq. \eqref{eq:5}, we have
\begin{equation}
\frac{\p\zeta}{\p x} + \nabla\times(\zeta\times\vg) = \nabla\times [- 2\bb{\Omega} \times\vg + 3\Omega ^2x\mathbf{e}_x] + \bb{S},
\label{eq:C4}
\end{equation}
and the source terms are
\begin{equation}
\bb{S} = \nabla\times \bigg[- \frac{\nabla P}{\rg}  + \frac{\epsilon}{\ts}(\vd-\vg)  + \nu\nabla^2\vg + F(x)\mathbf{e}_y \bigg].
\end{equation}
They correspond to baroclinity, drag force, gas viscosity, and forcing, respectively. By noting that 
\begin{equation}
\nabla\times(\zeta\times\vg) = (\vg\cdot\nabla)\zeta + \zeta(\nabla\cdot\vg) ,
\end{equation}
and 
\begin{equation}
\nabla\times [- 2\bb{\Omega} \times\vg + 3\Omega ^2x\mathbf{e}_x] = -2\bb{\Omega} (\nabla\cdot\vg),
\end{equation}
and that all terms in \ref{eq:C4} are in $z$-direction, we arrive at 
\begin{equation}
\frac{Dq}{Dt} = \frac{S}{\rg},
\end{equation}
where $D/Dt=\p/\p t + \vg\cdot\nabla$ is the material derivative.
Then, the perturbed vortensity equation is 
\begin{equation}
(-i\omega + i\ky v_\mathrm{gy0})\delta q + \delta\vgx \frac{\p q_0}{\p x} = S',
\end{equation}
where 
\begin{equation}
S'= S'_\mathrm{drag} + S'_\mathrm{visc} + S'_\mathrm{forc}.
\end{equation}
% and 
% \begin{equation}
% S'_\mathrm{bar} = \frac{i\ky}{\rho_{g0}^3}\bigg[-\frac{\p P_0}{\p x}\delta\rho+\frac{\p\rho_\mathrm{g0}}{\p x}\delta P \bigg],
% \end{equation}
Different from \citetalias{lb23}, the equation of state in this work is barotropic, and hence no baroclinic term involved in the perturbed vortensity equation. In practice, the viscosity and forcing terms are much smaller than the drag force term, and 
\begin{equation}
S'\approx S'_\mathrm{drag}= \frac{1}{\rg}\frac{\epsilon}{\ts}\nabla\times(\delta\vd-\delta\vg) .
\end{equation}

Following a similar derivation and utilizing equation \eqref{eq:4}, we obtain the dust vortensity equation,
\begin{equation}
\frac{Dq_\mathrm{d}}{Dt} = \frac{S_\mathrm{d}}{\rd},
\end{equation}
where 
\begin{equation}
\bb{S_\mathrm{d}} = \nabla\times \bigg[- \frac{1}{\ts}(\vd-\vg) + \frac{1}{\rd}\nabla\cdot(\rd\bb{v}_\mathrm{dif}\bb{v}_\mathrm{dif}) \bigg].
\end{equation}
The perturbed dust vortensity equation is 
\begin{equation}
(-i\omega + i\ky v_\mathrm{dy0})\delta q_\mathrm{d} + \delta\vdx \frac{\p q_\mathrm{d0}}{\p x} = S'_\mathrm{d}.
\end{equation}
The drag force term dominates over the dust diffusion term for $S'_\mathrm{d}$, 
\begin{equation}
S'_\mathrm{d}\approx S'_\mathrm{d,drag} = -\frac{1}{\rd\ts}\nabla\times(\delta\vd-\delta\vg) . 
\end{equation}

%%%%%%%%%%%%%%%%%%%%%%%%%%%%%%%%%%%%%%%%%%%%%%%%%%

\end{document}